\documentclass[aps,prd,preprint,showpacs,floatfix,preprintnumbers,nofootinbib,superscriptaddress,
showkeys]{revtex4}
\usepackage[utf8]{inputenc}
\usepackage{fancybox}
\usepackage{hhline}
\usepackage{dcolumn}
\usepackage{textcomp}
\usepackage{epsfig,graphics,graphicx}
\usepackage{amsfonts,amssymb,amsmath}
\usepackage{pifont}
\usepackage{bm}
\usepackage{longtable} 
\usepackage{appendix}
\usepackage{lscape}
\usepackage[mathscr]{euscript}
\usepackage{mathrsfs}
\usepackage{multirow}

\usepackage{rotating}
\usepackage{color}   

\newcommand{\beq}{\begin{equation}}
\newcommand{\eeq}{\end{equation}}
\newcommand{\bea}{\begin{eqnarray}}
\newcommand{\eea}{\end{eqnarray}}

\newcommand{\eps}{\epsilon}
\newcommand{\ord}[1]{{\cal{O}}( #1 )}
\newcommand{\B}{{\bf B}}
\newcommand{\Bdag}{{\bf B^\dagger}}
\newcommand{\p}{{\mathfrak{p}}^0}

\DeclareFontFamily{OT1}{pzc}{}
\DeclareFontShape{OT1}{pzc}{m}{it}%
              {<-> s * [0.900] pzcmi7t}{}
\DeclareMathAlphabet{\mathpzc}{OT1}{pzc}%
                                 {m}{it}

\DeclareMathAlphabet{\mathcalligra}{T1}{calligra}{m}{n}

\usepackage{hyperref}
\hypersetup{pdfauthor={me}, colorlinks=true, citecolor=blue, urlcolor=blue, linkcolor=black}
\begin{document}
\preprint{\vbox{\hbox{ JLAB-THY-12-1644} }}
\title{\phantom{x}
\vspace{-0.5cm}     }
\title{Baryon Masses and Axial Couplings  in the Combined  ${\mathbf{1/N_c}}$ and Chiral Expansions }

\author{A.~Calle~Cord\'on}\email{cordon@jlab.org}
\affiliation{Thomas Jefferson National Accelerator Facility, Newport News, Virginia 23606, USA.}
\author{J.~L.~Goity}\email{goity@jlab.org}
\affiliation{Thomas Jefferson National Accelerator Facility, Newport News, Virginia 23606, USA.}
\affiliation{Department of Physics, Hampton University, Hampton, VA 23668, USA.}
%
 
\begin{abstract}
The effective theory for baryons with  combined $1/N_c$ and chiral expansions is  analyzed for non-strange baryons.  Results for baryon masses and axial couplings are obtained in the small scale expansion, to be coined as the $\xi$-expansion, in which  the $1/N_c$ and the low energy power countings are linked according to $1/N_c=\ord{\xi}=\ord{p}$. Masses and axial couplings are analyzed to $\ord{\xi^3}$ and $\ord{\xi^2}$ respectively, which correspond to next-to-next to leading order evaluations, and require one-loop contributions in the effective theory.
 The spin-flavor approximate symmetry, consequence of the large $N_c$ limit in baryons,   plays a very important role in the real world with $N_c=3$ as shown by the analysis of its breaking in the masses and the axial couplings.  Applications to the recent lattice QCD results on baryon masses and the nucleon's axial coupling are presented. It is shown that those results are naturally described within the effective theory at the order considered in the $\xi$-expansion.
\end{abstract}


\maketitle

\tableofcontents


\section{Introduction}
\label{sec:Intro}

The low energy effective theory for baryons is a topic that has evolved over time through several approaches and improvements. The early version of baryon  Chiral Perturbation Theory (ChPT)
\cite{Pagels:1974se}
 evolved into  the various effective field theories based on effective chiral Lagrangians \cite{Weinberg:1968de,Coleman:1969sm,Callan:1969sn}, starting with the relativistic version 
\cite{Gasser:1987rb,Bernard:1995dp} 
or   Baryon ChPT (BChPT), followed by the non-relativistic version based in an expansion in the inverse baryon mass 
\cite{Jenkins:1990jv,Bernard:1995dp}
 or   Heavy Baryon ChPT (HBChPT), and  by  manifestly Lorentz covariant versions based on the IR regularization scheme \cite{Ellis:1997kc,Becher:1999he,Fuchs:2003qc}. 
  In all these versions of the baryon effective theory a consistent low energy expansion can be implemented. The most important issue,  which became apparent quite early,  was the convergence of the low energy expansion. Being an expansion that progresses in steps of $\ord{p}$ in contrast to the expansion in the pure Goldstone Boson sector where the steps are $\ord{p^2}$, it is natural to expect a slower rate of convergence. However, a key factor  with the convergence   has to do with the important effects due to the closeness in mass of the spin 3/2 baryons. It was realized \cite{Jenkins:1991es}, that the inclusion of those degrees of freedom play an important role in improving the convergence of the one-loop contributions to certain observables such as  the $\pi$-$N$ scattering amplitude and the axial currents and magnetic moments. There have been since then numerous works including spin 3/2 baryons \cite{Hemmert:1996xg,Hemmert:1997ye,Hemmert:2003cb,Fettes:2000bb,Procura:2006bj,Hacker:2005fh,Bernard:2003xf,Bernard:2005fy,Procura:2006gq}. The key enlightenment resulted from the study of baryons in the large $N_c$ limit of QCD \cite{'tHooft:1973jz}. It was realized that in that limit baryons behave very differently than mesons \cite{Witten:1979kh}, in particular because their masses scale like $\ord{N_c}$ and the $\pi$-baryon couplings are $\ord{\sqrt{N_c}}$. These properties were shown to require for consistency, that at large $N_c$ baryons must respect a dynamical contracted spin-flavor symmetry $SU(2 N_f)$, $N_f$ being the number of light flavors \cite{Gervais:1983wq,Gervais:1984rc,Dashen:1993as,Dashen:1993ac}, broken by effects ordered in powers of $1/N_c$ and in  the quark mass differences. The inclusion of the consistency requirements of the large $N_c$ limit into the effective theory came naturally through a combination of the $1/N_c$ expansion and  HBChPT 
\cite{Jenkins:1995gc},  which is the framework followed in the present work. The study of one-loop corrections in that framework 
 was first  carried out in Refs.~\cite{Jenkins:1995gc,FloresMendieta:1998ii,FloresMendieta:2006ei}.  In the  combined theory one has to deal with the fact that the $1/N_c$ and Chiral expansions do not commute \cite{Cohen:1992uy}. The reason is due to the presence of the baryon  mass splitting scale of $\ord{1/N_c}$ ($\Delta-N$ mass difference), for which it becomes necessary to specify its order in the low energy expansion. Thus the $1/N_c$ and Chiral expansions must be linked. Particular emphasis will be given to the specific linking in which the baryon mass splitting is taken to be $\ord{p}$ in the Chiral expansion, and which will be called the $\xi$-expansion. Following references   \cite{Jenkins:1995gc,FloresMendieta:1998ii,FloresMendieta:2006ei}, the theoretical framework is presented here in   detail, in particular the power countings, the renormalization,   and  the linked $1/N_c$ and low energy expansions, along with observations that further clarify the significance of the framework. 

The very significant contemporary progress in the calculations of baryon observables in lattice QCD   (LQCD) \cite{Hagler:2009ni,Fodor:2012gf,Alexandrou:2011iu} opens new opportunities for further understanding the low energy effective theory of baryons. The   determination of  the quark mass dependence of the various low energy observables, such as masses, axial couplings, magnetic moments, electromagnetic polarizabilities, etc.,  are of key importance as a significant  test of the effective theory, in particular  its range of validity in quark masses,  as well as for the determination of its low energy constants (LECs). Lattice results for the $N$ and $\Delta$ masses  \cite{Durr:2008zz,WalkerLoud:2008bp,Aoki:2008sm,Lin:2008pr,Alexandrou:2009qu,Aoki:2009ix,Aoki:2010dy,Bietenholz:2011qq}  and the axial coupling $g_A$ of the nucleon \cite{Edwards:2005ym,Bratt:2010jn,Alexandrou:2010hf,Yamazaki:2008py,Yamazaki:2009zq,Lin:2008uz} at varying quark masses are analyzed with  the purpose of testing   the effective theory presented here. This in turn  can give   insights on LQCD results, in particular an understanding on  the role and relevance of  including  the spin 3/2 baryons consistently with large $N_c$ requirements.

This work is organized as follows. In Section~\ref{sec:Framework} the framework for the combined   ${ { 1/N_c}}$ and HBChPT expansions is presented.
Section~\ref{sec:Masses} presents the evaluation  of the baryon masses  and Section~\ref{sec:Axial} the one for axial couplings at the one-loop level. Section~\ref{sec:Lattice} is devoted to applying  those  results  in the $\xi$-expansion  to LQCD results. Finally,   Section~\ref{sec:Conclusions} is devoted to   observations and  conclusions . Several appendices present useful material used in the calculations, namely,  
Appendix~\ref{sec:Algebra} on spin-flavor algebra, 
Appendix~\ref{sec:Symmetries} on symmetries, 
Appendix~\ref{sec:Tools} on the construction of effective Lagrangians, and 
Appendix~\ref{sec:ME} on useful matrix elements of spin-flavor operators.

\section{Framework for the  combined ${\mathbf{\rm 1/N_c}}$ expansion and Baryon Chiral Perturbation Theory}
\label{sec:Framework}

In this section the framework for the combined $1/N_c$ and chiral expansions in baryons is presented in some detail along similar lines as in the original works \cite{Jenkins:1995gc,FloresMendieta:1998ii,FloresMendieta:2006ei}. The symmetries that the effective Lagrangian must respect in the chiral and large $N_c$ limits are chiral $SU_L(N_f)\times SU_R(N_f)$ and contracted dynamical spin-flavor symmetry $SU(2N_f)$\cite{Gervais:1983wq,Gervais:1984rc,Dashen:1993ac,Dashen:1993as}~\footnote{See also Appendix~\ref{sec:Symmetries}.}. $N_f$ is the number of light flavors, and in this work $N_f=2$.  In the limit $N_c\to \infty$ the spin-flavor symmetry requires baryons to belong into degenerate multiplets of $SU(4)$. In particular, the  ground state (GS)  baryons belong into a symmetric $SU(4)$ multiplet,  which consists of states with  $I=S$, where  $S$ the baryon spin and $I$ its isospin. At finite $N_c$ the spin-flavor symmetry is broken by effects suppressed by powers of $1/N_c$,  and the baryon mass splittings  in the GS multiplet are proportional to $(S+1)/N_c$. The effects of finite $N_c$ are then implemented as an expansion in $1/N_c$ at the level of the effective Lagrangian.  Because baryon masses scale as proportional to $N_c$, it becomes natural to use the framework of HBChPT \cite{Jenkins:1990jv,Jenkins:1991ne}, where the expansion in inverse powers of the baryon mass becomes part of the $1/N_c$ expansion. The framework presented next  follows that of Refs. \cite{Jenkins:1995gc,FloresMendieta:1998ii}.

The non-relativistic baryon field, denoted  by $\bf B$, consists of the symmetric spin-flavor $SU(4)$ multiplet with states $I=S
$, $S=1/2,\cdots,N_c/2$ ($N_c$ odd). Chiral symmetry is realized in the usual non-linear way on $\bf B$, namely \cite{Weinberg:1968de,Coleman:1969sm,Callan:1969sn}:
\beq
(L,R):{\bf B}=h(L,R,u){\bf B},
\label{eq:LR-transf}
\eeq
where $L(R)$ is a $SU_{L(R)}(2)$ transformation,   $u$ is given in terms of the pion fields $\pi^a$ by  $u=\exp(i \pi^a I^a/F_\pi)$, where the isospin generators $I^a$ are normalized by the commutation relations $[I^a,I^b]=i\epsilon_{abc} I^c$,  $F_\pi=92.4$ MeV,  and $h(L,R,u)$ is an $SU_{I}(2)$ isospin transformation which in any representation of Isospin satisfies $R \,u \,h^\dag(L,R,u)=h(L,R,u) \,u\, L^\dag$. The chiral covariant derivative $D_\mu {\bf B}$ is given by:
\bea
D_\mu \B&=&\partial_\mu \B-i \Gamma_\mu  \B ,\nonumber\\
\Gamma_\mu&=&\frac{1}{2}\,(u^\dag(i\partial_\mu+r_\mu) u+u(i\partial_\mu+ l_\mu) u^\dag),
\label{eq:Dmu}
\eea
where   $l_\mu  = v_\mu - a_\mu$  and $r_\mu = v_\mu + a_\mu$   are gauge sources.  
Another necessary   building block of the effective chiral Lagrangian is the axial Maurer-Cartan one-form:
\beq
u_\mu=u^\dag(i\partial_\mu+r_\mu) u-u(i\partial_\mu+ l_\mu) u^\dag,~~~~(L,R):u_\mu=h(L,R,u)u_\mu h^\dagger(L,R,u).
\label{eq:Maurer-Cartan}
\eeq

For later use, the following notation will be used:   $\langle A\rangle\equiv{ \rm Tr }A$   for flavor traces, and the definition  $A^a\equiv \frac{1}{2} \langle \tau^a A\rangle$, where  $A$ is in the fundamental representation, which implies that in an arbitrary isospin representation  $A=2 A^a I^a$ (since in the fundamental representation,  $I^a=\tau^a/2$). The definition  $\tau^0=I_{2\times2}$ is used. 

Since $F_\pi=\ord{\sqrt{N_c}}$, $u$, $u_\mu$ and $\Gamma_\mu$ contain different orders in the expansion in powers of $1/N_c$. The contracted $SU(4)$ transformations (see Appendix~\ref{sec:Algebra}) are generated by $\{S^i,I^a, X^{ia}\}$, where $X^{ia}=G^{ia}/N_c$ are semiclassical at large $N_c$, i.e., commute with each other.  The ordering in $N_c$ of the matrix elements of the spin-flavor generators  in  states with $S=\ord{N_c^0}$ are as follows: $S^i=\ord{N_c^0}$,  $I^a=\ord{N_c^0}$,  and $G^{ia}=\ord{N_c}$.  While infinitesimal $SU(4)$ transformations generated by   $I^a$ correspond to the usual isospin transformations when acting on pions, the ones generated by $X^{ia}$   affect only the baryons (one can  define these generators  to not affect the pion field   as shown in  Appendix~\ref{sec:Symmetries}). The effective Lagrangian can be systematically written as a power series in the low energy expansion or Chiral expansion, and simultaneously in $1/N_c$. It is  most convenient to write the Lagrangian to be manifestly chiral invariant as is usually done. The low energy constants (LECs) will themselves admit an expansion in powers of $1/N_c$. For the HBChPT expansion the large mass of the expansion is taken to be the spin-flavor singlet component of the baryon masses, $M_0=N_c\, m_0$ ($m_0$ can be considered here to be a LEC defined in the chiral limit and which will have itself an expansion in $1/N_c$).  To $\ord{1/N_c}$ baryon masses will read \cite{Dashen:1993ac,Dashen:1993as}:
\beq
m_\B(S)=M_0+\frac{C_{HF}}{N_c} S(S+1)+c_1 \,N_c\,M_\pi^2+\cdots.
\label{eq:LO-mB}
\eeq
In the following we will define 
\beq
\delta m(S)\equiv \frac{C_{HF}}{N_c} S(S+1)+c_1 \,N_c\,M_\pi^2,
\label{eq:dm}
\eeq
 which will be useful in the implementation of the expansion discussed later.
The baryon mass splittings due to the hyperfine term, second term in   Eq.~\eqref{eq:LO-mB}, must be considered to be a small energy scale. It becomes necessary to establish of what order that term  is in the low energy expansion, as it naturally appears in combinations with powers of $M_\pi$ when loop diagrams are calculated. This fact  implies that the low energy and $1/N_c$ expansions do not  commute \cite{Cohen:1992uy,Cohen:1996zz}, and the natural way to proceed is therefore  to link the two expansions. For the purpose of organizing the effective Lagrangian it is convenient to establish  the link between the two expansions.
 In the real world with $N_c=3$ the $\Delta-N$ mass splitting is about 300 MeV, and  therefore  it is reasonable to count that quantity as $\ord{p}$ in the low energy expansion: the expansion where $1/N_c=\ord{p}=\ord{\xi}$ will be adopted in  what follows, and it will be called $\xi$-expansion.  This power counting corresponds to the so called small scale expansion (SSE) \cite{Hemmert:1997ye},  now consistently implemented in the context of the $1/N_c$  expansion.  Whenever appropriate,  it will be indicated which aspects of the analysis are general and which are only valid in that expansion. Up to $\ord{\xi}$ the baryon effective Lagrangian reads \cite{Jenkins:1995gc}:
 \bea
{ \cal{L}}_\B^{(1)}&=&\Bdag\left(i D_0 +  \mathring{g}_A
 u^{ia}G^{ia}-\frac{C_{HF}}{N_c}{ \vec{S}^2}-\frac{c_1}{2} N_c\; \chi_+\right)\B,
\label{eq:Lagrangian-LO}
 \eea
where   $\mathring{g}_A$ is the axial coupling in the chiral and large $N_c$ limits (it has to be rescaled by a factor 5/6 to coincide with the usual axial coupling as defined for the nucleon), $\chi_+$ is the source containing the quark masses: specifically $\chi_+=2 M_\pi^2+\cdots$ (see Appendix~\ref{sec:Tools} ). Here one notes an important point  which will be present in other instances as well: the baryon mass dependence on the current quark mass behaves at $\ord{N_c\; M_\pi^2}$ ($c_1$ is of zeroth order in $N_c$), and this indicates that in a strict large $N_c$ limit the expansion in the quark masses of certain quantities such as the baryon masses cannot be defined due to divergent coefficients of $\ord{N_c}$. 

The Lagrangian is  manifestly  invariant under  chiral transformations,  translations and rotations (the latter also involving obviously the action of the $S^i$ generators of $SU(4)$). 
Under an infinitesimal transformation generated by the spin-flavor generators $X^{ia}$, the Lagrangian  \eqref{eq:Lagrangian-LO} is transformed according to:
\beq
\delta { \cal{L}}_\B^{(1)}=-i \;\delta\alpha^{ia}\; [X^{ia},{ \cal{L}}_\B^{(1)}].
\eeq
According to this, and  using the commutation relations in Appendix \ref{sec:Algebra}, the kinetic term changes by terms  $\ord{1/N_c^2}$, the term proportional to   $\mathring{g}_A$, which contains the $\pi \B\B'$ interaction  and   the leading order terms of the axial currents, changes by terms which are a factor $\ord{1/N_c^2}$ smaller than the original term, and the term proportional to $c_1$, which gives the leading order (LO)    $\sigma$-term in the baryon masses,  is a spin-flavor singlet and thus invariant under spin-flavor transformations. 
 Finally, the hyperfine term proportional to $C_{HF}$ is the one providing the dominant spin-flavor symmetry breaking effects, because it is modified by  terms $\ord{1/N_c}$, which is the same order as the hyperfine term itself (this is so because $[{\vec{S}^2},X^{ia}]=\ord{N_c^0}$). 
   The construction of higher order Lagrangians can be accomplished using the tools provided in Appendix~\ref{sec:Tools}. 
   
   The operators appearing in the effective Lagrangian are normalized in such a way that all the LECs are of zeroth order in $N_c$. Therefore, the $1/N_c$ power  of a  Lagrangian term  with $n_\pi$ pion fields is given by \cite{Dashen:1994qi}: 
\beq
n-1-\kappa+\frac{n_\pi}{2},
\label{eq:N-counting}
\eeq 
where the spin-flavor operator is $n$-body ($n$ is the number of factors of $SU(4)$ generators appearing in the operator),  and $\kappa$ is basically the number of factors of the generators $G^{ia}$ remaining after reducing the operator using commutators.   The last term, ${n_\pi}/{2}$, stems from the factor $(1/F_\pi)^{n_\pi}$  carried by any term with $n_\pi$ pion fields. It is opportune to point out that commutators of spin-flavor operators will always reduce the $n$-bodyness of the product of operators: e.g., let ${\cal{G}}$ be any generator of $SU(4)$, and consider the commutator $[{\cal{G}},\vec{S}^2]=\{S^i,[{\cal{G}},S^i]\}$. In principle this looks like a three-body operator, but because  $[{\cal{G}},S^i]$ is a 1-body operator, $[{\cal{G}},\vec{S}^2]$ is actually a 2-body operator. 

\subsection{Consistency of the $1/N_c$ expansion}
\label{sec:Consistency}

The consistency of the $1/N_c$ expansion in QCD gives rise to the   dynamical spin-flavor contracted $SU(2N_f)$ symmetry in baryons at large $N_c$.  At the baryon level that symmetry can be deduced as the result of consistency  or correct $N_c$ power counting of observables in which pion-baryon couplings are involved. This is  because the pion-baryon coupling is $\ord{\sqrt{N_c}}$   from Witten's counting rules~\cite{Witten:1979kh}. In particular the consistency of pion-baryon scattering is a direct way of deriving the existence of the dynamical spin-flavor symmetry \cite{Dashen:1993ac,Dashen:1993as}. In general, for any quantity  there must be cancellations between the  terms with the \lq\lq wrong\rq\rq power counting stemming from different Feynman diagrams.
For instance, baryon masses are $\ord{N_c}$, and therefore pion loop contributions cannot give contributions which scale with a higher power of $N_c$. On the other hand, the baryon mass splittings are $\ord{1/N_c}$, and loop contributions must respect that scaling. Similarly, in the axial currents, whose matrix elements are $\ord{N_c}$ such cancellations occur when loop corrections are calculated. All this will be illustrated in the application to baryon masses and axial couplings   discussed later. Although certain key cancellations must be exact in the large $N_c$ limit, the analysis of LQCD  results will show that they are very significant   in the physical world where  $N_c=3$.

\subsection{$\xi$ power counting}
\label{sec:power counting}

The terms in the effective Lagrangian  are constrained in their  $N_c$ dependence by the requirement of the consistency of QCD at large $N_c$. This     constraint  is in the form of a lower bound in the power in $1/N_c$ for each term one could write down in the Lagrangian.     This leads   to   constraints  on the $N_c$ dependencies of the ultra-violet (UV) divergencies,  which have to be subtracted by the corresponding counter-terms in the Lagrangian. 
One very important point to mention is that the UV divergencies are necessarily polynomials in low momenta $p$ (derivatives), in $M_\pi^2$ and in $1/N_c$ (modulo factors of $1/\sqrt{N_c}$ due to   $1/F_\pi$ factors in terms where pions are attached). Therefore, the structure of counter-terms is independent of any linking between the $1/N_c$ and chiral expansions. For this reason, one  can simply take the large $N_c$ and low energy limits independently in order to determine the UV divergencies. 
For a connected diagram with $n_B$ external baryon legs, $n_\pi$ external pion legs, $n_i$ vertices of type $i$ which has $n_{B_i}$ baryon legs and $n_{\pi_i}$ pion legs, and $L$ loops, the following topological relations hold \cite{Weinberg:1995mt,Weinberg:1991um}:
\bea
L=1+I_\pi+I_B-\sum n_i ,~~~~2 I_B+n_B=\sum n_i \;n_{B_i},~~~~2 I_\pi+n_\pi=\sum n_i \;n_{\pi_i},
\label{eq:topol}
\eea
where $I_\pi$ is the number of pion propagators and $I_B$ the number of baryon propagators.  

The chiral or low energy    order of a diagram, where  $\nu_{p_i}$ is the   chiral power of the vertex of type $i$, is   then given by \cite{Weinberg:1991um}:
\beq
\nu_p = 2 - \frac{n_B}{2} + 2 L +\sum_i n_i\;(\nu_{p_i} + \frac{n_{B_i}}{2} -2),
\label{eq:chiral-counting}
\eeq
Note that  $n_{B_i}$ is equal to 0 or 2 in the single baryon sector.
 
 On the other hand, the $1/N_c$ power of a connected diagram is determined by looking only at the vertices: the order in $1/N_c$ of a vertex of type $i$ is given according to Eq. \eqref{eq:N-counting} by: $\nu_{O_i}+\frac{n_{\pi_i}}{2}$, where $\nu_{O_i}$ is the order of the spin-flavor operator. Thus, the $1/N_c$ power of a diagram, upon use of the third Eq. \eqref{eq:topol}, is given by:
\beq
\nu_{1/N_c} =\frac{n_\pi}{2} +I_\pi+\sum{n_i\; \nu_{O_i}}~,
\label{eq:Nc-counting}
\eeq
where $n_\pi$ is the number of external pions, and $\nu_{O_i}$ the $1/N_c$ order of the spin-flavor operator of the  vertex of type $i$. Since $\nu_{O_i}$ can be negative (due to factors of $G^{ia}$ in vertices), one can think of individual diagrams with  $\nu_{1/N_c}$ negative and violating large $N_c$ consistency, requiring cancellation with other diagrams. Such a sum will have to respect the   mentioned lower bound on the $1/N_c$ power corresponding to the sum of such diagrams. The explicit example of such cancellation in the axial currents at one-loop  is given in Section IV. 
 
One can determine now the nominal counting of the one-loop contributions to the baryon masses and axial currents. The LO baryon masses are $\ord{N_c}$, Eq. \eqref{eq:LO-mB}.The one loop correction shown in Fig.~\ref{fig:self-energy} has: $(L=1,~ n_B=2,~ n_\pi=0, ~n_1=2,~ \nu_{O_1}=-1,~n_{B_1}=2,~\nu_{p_1}=1)$ giving $\nu_p=3$ as it is well known, and  $\nu_{1/N_c}=-1$. Since there is only one possible diagram, this must be consistent by contributing $\ord{N_c}$ to the spin-flavor singlet component of the masses, which is the case as shown in the next section.  For the axial currents one has the diagrams in Fig.~\ref{fig:1-loop-vertex}. The current at tree level is $\ord{N_c}$, and the sum of the diagrams cannot scale like a higher power of $N_c$. Performing the counting for the individual diagrams one obtains:
$\nu_p(j)=2$ for $j=1,\cdots,4$, and $\nu_{1/N_c}(j)=-2$, $j=1,2,3$ and $\nu_{1/N_c}(4)=0$. Thus a cancellation must occur of the $\ord{N_c^2}$ terms when the contributions to the axial currents by diagrams 1, 2 and 3 are added. Since the acceptable bound is that the sum be $\ord{N_c}$, one concludes that the axial current has, at one-loop,  corrections $\ord{p^2 N_c}$ or higher.

One can consider the case of two-loop diagrams, in particular  diagrams where the same pion-baryon  vertex Eq.(\ref{eq:Lagrangian-LO}) appears four times.  For the masses one has $\nu_p(j)=5$, and individual diagrams give $\nu_{1/N_c}=-2$. A cancellation must occur to restore the bound on the $N_c$ counting for the masses, i.e., $\ord{N_c}$. Thus, at two-loops the UV divergencies of the masses must be $\ord{p^5 N_c}$ or higher. For the axial currents a similar discussion requires that  counter-terms to the axial currents must be $\ord{p^4 N_c}$ or higher.

Defining the linked power counting $\xi$ by: $\ord{1/N_c}=\ord{p}=\ord{\xi}$,   the  $\xi$ order of a given Feynman diagram will be simply equal to  $\nu_p+\nu_{1/N_c}$ as given by Eqs.\eqref{eq:chiral-counting} and  \eqref{eq:Nc-counting}, which upon use of the topological formulas Eq.\eqref{eq:topol} leads to:
\beq
\nu_\xi=1+3 L+\frac{n_\pi}{2} +\sum_i n_i\;(\nu_{O_i}+\nu_{p_i}-1).
\label{eq:xi-counting}
\eeq
  The $\xi$-power counting of the UV divergencies is obvious from the earlier discussion. At one-loop one finds  that the masses have $\ord{\xi^2}$ and $\ord{\xi^3}$ counter-terms, while the axial currents will have $\ord{\xi}$ and $\ord{\xi^2}$ counter-terms. To two loops one expects  $\ord{\xi^4}$ and $\ord{\xi^5}$, and  $\ord{\xi^3}$ and $\ord{\xi^4}$ counter-terms for masses and axial currents respectively. The non-commutativity of limits is manifested in the finite terms where $M_\pi$ and or momenta  and $\delta m$ appear combined in non-analytic terms, and are therefore sensitive to the linking of the two expansions.

\section{Baryon masses}
\label{sec:Masses}

\begin{center}
\begin{figure}[h]
\centerline{\includegraphics[width=8.cm,angle=-0]{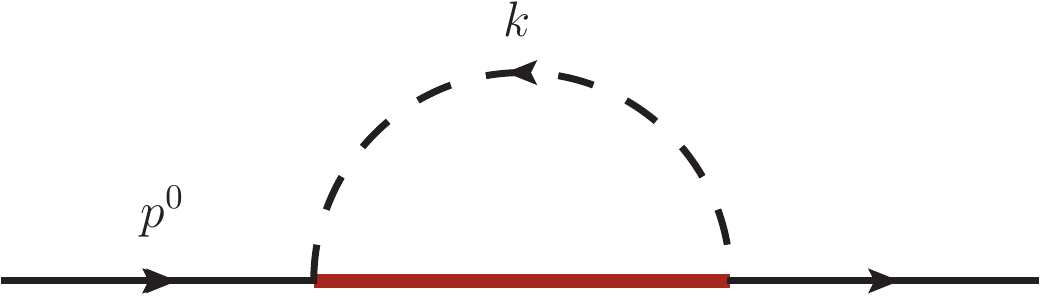}}
\caption{One-loop contribution to baryon self energy. The thick propagator indicates sum over all possible baryons that can contribute.}
\label{fig:self-energy}
\end{figure}
\end{center}

In this section baryon masses are analyzed to order $\xi^3$, or next-to-next to leading order  (NNLO), in the limit of exact isospin symmetry. To that order  the   mass of the baryon of spin $S$ reads:
\beq
m_\B(S)=N_c m_0+\frac{C_{HF}}{N_c} S(S+1)+c_1 N_c M_\pi^2+\delta m^{1-loop+CT}_\B(S),
\label{eq:baryon-mass}
\eeq
where $\delta m^{1-loop+CT}_\B(S)$ involves contributions from the one-loop diagram in Fig.~\ref{fig:self-energy}, and CT  denotes  counter-terms. From both types of contributions, there are $\ord{\xi^2}$ and $\ord{\xi^3}$ terms,  and the calculation is exact at the  latter order, as can be deduced from the previous discussion on power counting. 
 Notice that $C_{HF}$ is equal to the LO term in $M_\Delta-M_N$ in the real world $N_c=3$.

\vspace{-1cm}
The leading 1-loop correction to the baryon self energy, diagram in Fig.~\ref{fig:self-energy},  can be calculated through the matrix element $\langle \B\mid  \delta\Sigma_{1-loop}\mid \B \rangle$, with:
 
\bea
 \delta\Sigma_{1-loop}&=&i\,\frac{ \mathring{g}_A^2}{F_\pi^2}\;\frac{1}{d-1}\;\sum_{n} G^{ia} {\cal{P}}_n
G^{ia}\;I_{1-loop}(\delta m_n-p^0,M_\pi)\, , 
\label{eq:delta-Sigma}
\eea
where $n$ indicates the possible intermediate baryon spin-isospin states in the loop,  ${\cal{P}}_n$ are the corresponding spin-flavor projection operators, $\delta m_n=\delta m(S_n)$, and  the loop integral is calculated in dimensional regularization with the result,
\bea
\label{eq:I1-loop}
 I_{1-loop}(Q,M_\pi) &=& \int \frac{d^d k}{(2 \pi)^d}\;
\frac{\vec{k}^2}{k^2-M_\pi^2+i\eps}\; \frac{1}{k^0-Q + i\eps}\nonumber\\
&=& \frac{i}{16\pi^2}\left\{ Q\;\left((3M_\pi^2-2Q^2)(\lambda_\eps-\log\frac{M_\pi^2}{\mu^2})+(5 M_\pi^2-4 Q^2)\right)\right.\nonumber\\
&+& 2\pi(M_\pi^2-Q^2)^{3/2} 
+\left. 4   (Q^2-M_\pi^2)^{3/2} 
\tanh^{-1}\frac{Q}{\sqrt{Q^2-M_\pi^2}}\right\} 
\, ,
\eea
where $Q=\delta m_n-p^0$, $\lambda_\eps=\frac{1}{\eps}-\gamma+\log 4\pi$, and $\mu$ is the renormalization scale which will be taken later to be of the order of  $m_\rho$.  For the specific evaluation of $\delta\Sigma_{1-loop}$ for a given  baryon state denoted by $in$, $p^0=\delta m_{in}-\p$, where $\p$ is a residual energy  (when evaluated on an on-shell baryon it is the kinetic energy which is $\ord{p^2/N_c}$). The non-commutativity of the $1/N_c$ and  Chiral expansions of course resides in the non-analytic terms of the loop integral  through their dependence  on the ratio  $Q^2/M_\pi^2$. Notice that when the one loop integrals are written in terms of the residual momentum $\p$, they do not depend on the spin-flavor singlet piece of $\delta m$, namely the $\sigma$-term in Eq.\eqref{eq:dm}.

Appendix~\ref{sec:ME}  provides all the necessary  elements  for the evaluation of the spin-flavor matrix elements in Eq.~\eqref{eq:delta-Sigma} as well as in the calculation of the one-loop corrections to the axial currents below. The explicit final expressions for the self energy are not given here because they are too lengthy, but with those elements the reader can  easily obtain them.

The one-loop contribution to the wave function renormalization constant is given by:
\beq
\delta Z_{1-loop}=\left.\frac{\partial}{\partial \p}\delta\Sigma_{1-loop}\right\arrowvert_{{\p\to 0}}.
\label{eq:WF-ren-Z}
\eeq

The explicit evaluation of the ultraviolet divergent pieces of the self energy gives:
\bea
&&\delta\Sigma_{1-loop}^{UV}
=\frac{\lambda_\eps}{16 \pi^2}\frac{\mathring{g}_A
^2}{F_\pi^2}
 \\
&\times&\left\{\frac{C_{HF}}{24N_c}\left(-3M_\pi^2(3 N_c(4+N_c)-20 \vec{S}^2)+8\frac{C_{HF}^2}{N_c^2}(N_c(4+N_c)(3+5 \vec{S}^2)-4\vec{S}^2(6+7\vec{S}^2))\right)\right.\nonumber\\
&+&\left.\p\left(\frac{M_\pi^2}{2}(\frac{3}{8}N_c(4+N_c)-\vec{S}^2)-\frac{C_{HF}^2}{4N_c^2}(N_c(4+N_c)(3+2\vec{S}^2)-8\vec{S}^2(3+\vec{S}^2))\right)+\ord{\p{^{^2}}}\right\}.\nonumber
\eea
The UV divergent pieces start at $\ord{\xi^2}$. Note that the  UV divergencies in the mass (term independent of $\p$) is produced by the contribution of the partner baryon and is proportional to the mass splitting.  As is well known,  they are absent in HBChPT without explicit $\Delta$.  The $\ord{N_c^0}$ UV divergence is spin-flavor singlet and proportional to $M_\pi^2$,  while the contributions to mass splittings are  $\ord{1/N_c^2}$. 
Notice that the    leading  UV divergence of $\delta Z_{1-loop}$ is $\ord{M_\pi^2 N_c}$:  this is necessary as shown later for rendering the one-loop calculation of the axial currents consistent in the large $N_c$ limit. 
 Since the calculation is accurate to $\ord{\xi^3}$, additional terms in the effective Lagrangian up to that order are necessary for renormalization. The terms necessary for renormalizing the self energy are therefore the following:
\bea
{\cal{L}}_{\Sigma}^{CT}&=&\Bdag\left\{\frac{m_1(N_c)}{N_c} +\frac{C_{HF1}(N_c)}{N_c^2}\vec{S}^2+\frac{C_{HF2}(N_c)}{N_c^3}\vec{S}^4 \right.+\mu_1(N_c) \chi_+ +\frac{\mu_2(N_c)}{N_c} \chi_+ \vec{S}^2  \\
&+&\left. \left(\frac{w_1(N_c)}{N_c}+\frac{w_2(N_c)}{N_c}\vec{S}^2+\frac{w_3(N_c)}{N_c^3}\vec{S}^4+(z_1(N_c) N_c+\frac{z_2(N_c)}{N_c}\vec{S}^2)\chi_+\right)(iD_0-\delta m)\right\}\B,\nonumber 
\label{eq:self-energy-CT-lagrangian}
\eea
where    the residual energy $\p$ has been identified with the operator $(iD_0-\delta m)$. All LECs are here of the form $X(N_c)=X_0+X_1/N_c+\cdots$. 
Writing $X=X(\mu)+\gamma_{_X} \lambda_\eps$, one renormalizes the self energy to $\ord{\xi^3}$. The coefficients $\gamma_{_X}$  are determined from $\delta\Sigma_{1-loop}^{UV}$ given above. While the counter-terms are   defined such that    $X(\mu)$ is $\ord{N_c^0}$, it is possible that  $\gamma_{_X}$   is of higher order in $1/N_c$.
Notice that among the higher order terms there are terms which can be simply absorbed into $1/N_c$ corrections to the LECs of the lowest order Lagrangian, and into $m_0$. 

Finally, the baryon masses are given by:
\beq
m_\B=\langle \B\mid N_c {m}_0+\frac{C_{HF}}{N_c} \vec{S}^2+c_1 N_c M_\pi^2+(\delta\Sigma^{UV~finite}_{1-loop}+\delta\Sigma^{CT})\arrowvert_{\p=0}\;(1+
\delta Z^{UV~finite}_{1-loop}+\delta Z^{CT})\mid\B\rangle.
\label{eq:baryon-mass-1loop}
\eeq
Note that the correction to the wave function renormalization factor enters in the expression for the mass corrections: this is because $\delta\Sigma(\p=0)$ starts with  terms $\ord{\xi^2}$ and $\delta Z$ starts at $\ord{\xi}$,  therefore the $\ord{\xi^3}$  terms of the mass correction involve these lower order terms of the wave function renormalization.

The one-loop corrections and corresponding counter-terms contribute to the masses at $\ord{\xi^2}$ and $\ord{\xi^3}$, while in a strict large $N_c$ limit the following ordering is found:
\bea
M_B&=&\ord{N_c}+\ord{N_c M_\pi^2}+\ord{N_c^0 M_\pi}+\cdots~~,\nonumber\\
M_B-M_B' &=& \ord{\frac{1}{N_c}}+\ord{\frac{1}{M_\pi N_c^2}}.
\label{eq:mass-power-counting}
\eea
Obviously the term $\ord{\frac{1}{M_\pi N_c^2}}$ stems from the $1/N_c$ expansion of non-analytic terms and shows the non-commutativity of limits.

 The one loop correction with the vertex  proportional to $c_1$ in Eq.(\ref{eq:Lagrangian-LO}) gives $\ord{\xi^4}$ contributions to the masses, and is therefore beyond the accuracy considered here.

The $\sigma$-terms for $N$ and $\Delta$, defined by $\sigma_B=\hat{m}\frac{\partial m_B}{\partial \hat{m}}$ ($\hat{m}=\frac{1}{2}(m_u+m_d)$), 
are $\ord{\xi}$ with the   one-loop corrections contributing up to $\ord{\xi^3}$. The difference $\sigma^B-\sigma^{B'}$ is $\ord{\xi^2}$ and at that order it receives only finite contributions from the loop. This implies that the slopes of the $N$ and $\Delta$  masses as functions of $M_\pi$ are the same up to $\ord{\xi^2}$ deviations. This seems to be closely followed by the lattice QCD results analyzed later.
In the large $N_c$ limit, obviously $\sigma=\ord{N_c}$. The terms of that order are necessarily spin-flavor singlet, and taking the limit at fixed $M_\pi$ one finds $\sigma_\Delta- \sigma_N=\ord{1/N_c^2}$, a result similar to the one in the $\xi$-expansion.

\section{Axial couplings}
\label{sec:Axial}

In this section the evaluation of the axial couplings including corrections $\ord{\xi^2}$ is presented.  At that order the one-loop corrections must be calculated. 

 The matrix elements of interest for the axial currents are $\langle\B'\mid A^{ia}\mid\B\rangle$ evaluated at vanishing external 3-momentum. The axial couplings  are then defined by:
 \beq
 \langle\B'\mid A^{ia}\mid\B\rangle=g_A^{\B\B'} \; \frac{5}{6}\;{ \langle \B' \mid G^{ia} \mid \B \rangle} \, .
\label{eq:gA-formula}
 \eeq
The axial   couplings   defined here are $\ord{N_c^0}$. The $\ord{N_c}$ of the matrix elements of the axial currents is due to the operator $G^{ia}$. The factor $5/6$ mentioned earlier is included so  that $g_A^{NN}$ at $N_c=3$ exactly corresponds to the usual nucleon $g_A$, which has the value $1.2701 \pm 0.0025$ \cite{Beringer:1900zz}.   This definition of the axial couplings is convenient  in the context of the $1/N_c$ expansion, as the differences between the different axial couplings are $\ord{1/N_c^2}$.

\begin{center}
\begin{figure}[h]
\centerline{
\includegraphics[width=4cm,angle=0]{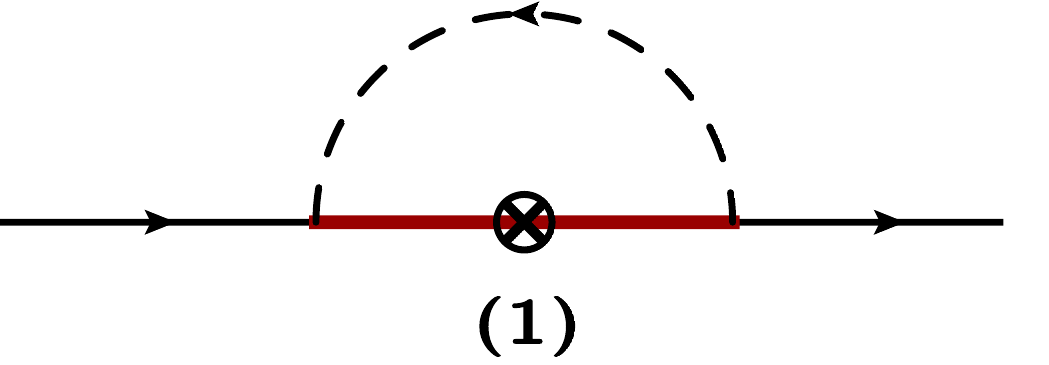}
\includegraphics[width=4cm,angle=0]{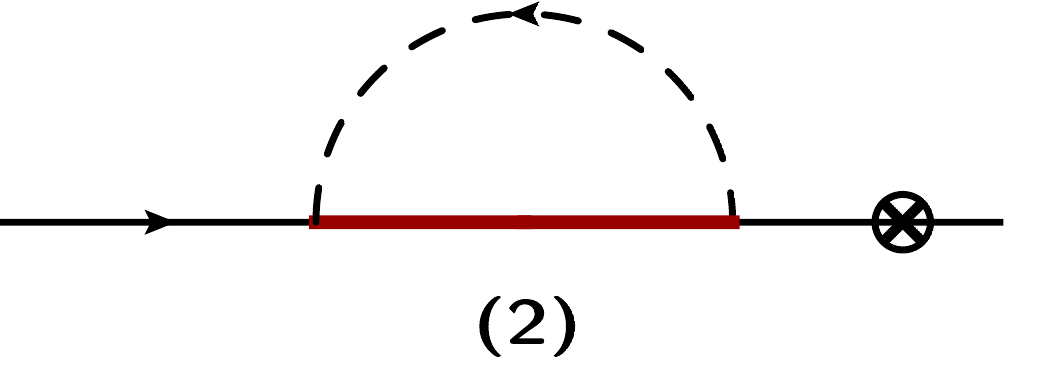}
\includegraphics[width=4cm,angle=0]{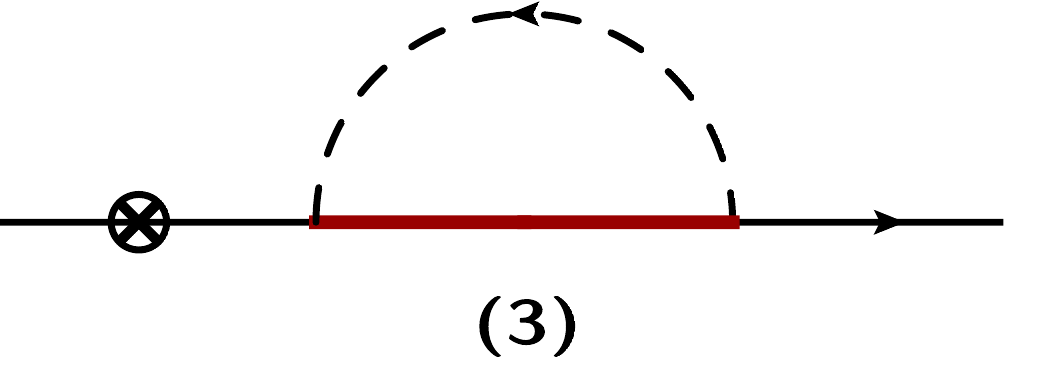}
\includegraphics[width=2cm,angle=0]{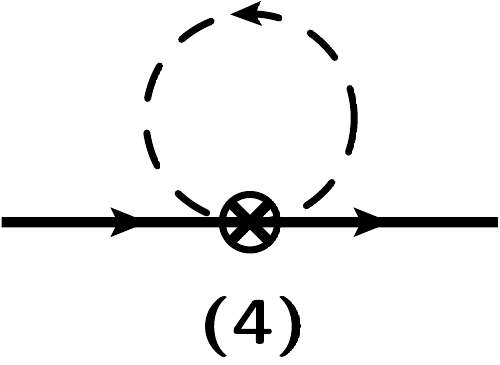}
}
\caption{Diagrams contributing to the 1-loop corrections to the axial-currents.   The crossed circle denotes   the axial-current operator.  }
\label{fig:1-loop-vertex}
\end{figure}
\end{center}

%
\vspace{-1cm}
The determination of the axial couplings to $\ord{\xi^2}$ require the calculation of the 1-loop corrections to the axial current. Only the contributions with no pion pole are necessary, and they are given by the diagrams  in Fig.~\ref{fig:1-loop-vertex}.  The resulting 1-loop contribution to the axial currents reads:
\bea
\delta A^{ia}_{1-loop} = \delta A^{ia}_{1-loop}(1) +  \delta A^{ia}_{1-loop}(2+3) + \delta A^{ia}_{1-loop}(4)\, ,
\label{eq:deltaA-1loop}
\eea
where $ \delta A^{ia}_{1-loop}(2+3)$ is given by a factor 1/2 times the no-baryon-pole contributions of diagrams (2+3).  The different contributions read as follows, where one needs to take the limits   $\p,{\mathfrak{p}'}^0\to 0$:
\bea
\delta A^{ia}_{1-loop}(1)&=&{-} i \frac{  \mathring{g}_A^3}{F_\pi^2}\frac{1}{d-1}\nonumber\\ &\times&\sum_{n,n'}G^{jb}{\cal{P}}_{n'}G^{ia}{\cal{P}}_nG^{jb}
\frac{I_{1-loop}(\delta m_n-p^0,M_\pi)-I_{1-loop}(\delta m_{n'}-p'^0,M_\pi)}{p^0-p'^0-\delta m_n+\delta m_{n'}},\nonumber\\
\delta A^{ia}_{1-loop}(2+3)&=&   \frac{ \mathring{g}_A}{2}
\left(G^{ia} \delta Z_{1-loop}+\delta Z_{1-loop} G^{ia}\right),\nonumber\\
\delta A^{ia}_{1-loop}(4)&=&-\frac{ \mathring{g}_A}{{  3} F_\pi^2}\Delta(M_\pi) G^{ia}.
\label{eq:A-1loop-contr}
\eea
Obviously, $G^{ia}$ and $\delta Z_{1-loop}$ do not commute in general. The pion tadpole integral in the last term is given by:
\beq
\Delta(M_\pi)=
-\frac{M_\pi^2}{16 \pi^2} (\lambda_\eps- \log\frac{M_\pi^2}{\mu^2}).
\label{eq:tadpole-integral}
\eeq
 Notice that  the contribution by diagram (4) is actually $\ord{\xi^4}$, and thus beyond the degree of accuracy of the present calculation. It can serve however as a measure of the size of the NNNLO corrections.

The corrections to the axial currents must scale as $\ord{N_c^\nu}$ with $\nu\leq 1$. While diagram (4) is $\ord{N_c^0}$ and therefore consistent in itself,   diagrams (1) and (2+3) above are $\ord{N_c^2}$. As shown in Ref. \cite{FloresMendieta:1998ii}, the offending terms cancel upon adding the diagrams. To test the cancellation it is sufficient to take the large $N_c$ limit at fixed $M_\pi$. A straightforward   evaluation leads to:
\beq
\left(\delta A^{ia}_{1-loop}(1)+\delta A^{ia}_{1-loop}(2+3)\right) {   \Big\vert_{N_c\to\infty}}  =-i\frac{  \mathring{g}_A^3}{F_\pi^2}\frac{1}{d-1}\left\lbrace\frac{1}{2}[[G^{jb},G^{ia}],G^{jb}]\frac{\partial}{\partial{\p}}I_{1-loop}(\p,M_\pi) + \cdots\right\rbrace \, ,
\label{eq:gA-cancel}
\eeq
where $\cdots$ indicate further  terms which are consistent with the $N_c$ power counting.   
The suppression of the $\ord{N_c^2}$ terms is direct consequence of the appearance of the commutator of two generators $G$, which is $\ord{N_c^0}$, when the diagrams are added up. In consequence the displayed terms are 
 $\ord{N_c^0}$.

The UV divergent contributions of the individual diagrams read:
\bea
\delta A^{ia}_{1-loop}(1)^{UV}&=& {-}
\frac{\lambda_\eps}{48 \pi^2}
 \frac{\mathring{g}_A^3}{F_\pi^2}
\Big\{ 3 M_\pi^2 G^{jb} G^{ia} G^{jb}\nonumber\\
&-& 2  \left(  \frac{C_{HF}}{N_c} \right)^2 \Big( G^{jb} G^{ia}
\left[\left[G^{jb},\vec{S}^2\right],\vec{S}^2\right]+\left[\vec{S}^2,\left[\vec{S}^2,G^{jb}\right]\right]G^{ia}G^{jb}  \nonumber\\
&+&\left[\vec{S}^2,G^{jb}\right]G^{ia}\left[G^{jb},\vec{S}^2\right] \Big)\Big\},\nonumber\\
\delta A^{ia}_{1-loop}(2+3)^{UV}&=&\frac{ \lambda_\eps}{96 \pi^2}
 \frac{\mathring{g}_A
^3}{F_\pi^2}
\Big\{ 3 M_\pi^2\{G^{ia},G^2\}
\nonumber\\
&-&2\left(\frac{C_{HF}}{N_c}\right)^2 \Big(G^{ia}G^{jb}[[G^{jb},\vec{S}^2],\vec{S}^2]+[\vec{S}^2,[\vec{S}^2,G^{jb}]G^{jb}G^{ia}\Big)\Big\},\nonumber\\
\delta A^{ia}_{1-loop}(4)^{UV}&=&   
\frac{\lambda_\eps \,
}{{ 48 \pi^2}}\frac{\mathring{g}_A}{ F_\pi^2 }\; M_\pi^2\; G^{ia}.
\label{eq:deltaA-1loop-detail}
\eea
One notices that only the terms proportional to $M_\pi^2$ in diagrams (1) and (2+3) diverge as proportional to $N_c^2$, while the terms proportional to $C_{HF}^2$ are $\ord{N_c^0}$.  Thus, only the $\ord{N_c^2}$  terms proportional to $M_\pi^2$ need  to be cancelled to give consistency.  One can easily check that such a  cancellation indeed occurs, leaving only terms $\ord{N_c^0}$. 
An explicit evaluation of these UV divergent terms   using the results from Appendix~\ref{sec:ME}   finally gives:
\bea
\delta {A^{ia}_{1-loop}}^{UV}&=&
\frac{\lambda_\eps}{32 \pi^2}
 \frac{\mathring{g}_A
}{F_\pi^2 N_c^2}\Big\{   (\frac{2}{3}+\mathring{g}_A^2) M_\pi^2 N_c^2 G^{ia}\nonumber\\
&+&\frac{C_{HF}^2 \mathring{g}_A^2}{3}  \Big(4-2 N_c(4+N_c)G^{ia}
  -7[ \vec{S}^2,[\vec{S}^2,G^{ia}]]
 + 4 \{ \vec{S}^2,G^{ia}\} \Big)  \Big\}.
 \label{eq:deltaA-1loop-UV}
 \eea
 
The terms in the Lagrangian needed to renormalize the axial currents are then the following:
\beq
{\cal{L}}{  _{A}^{CT}}=\Bdag u^{ia}\left( \frac{C^A_0}{N_c} G^{ia}+\frac{C^A_1}{4}\{\chi_+,G^{ia}\}+ \frac{C^A_2}{N_c^2} \{\vec{S}^2,G^{ia}\}+ \frac{C^A_3}{N_c} [\vec{S}^2,G^{ia}]+
\frac{C^A_4}{N_c} S^iI^a
\right)\B~.
\label{eq:axial-current-CT-lagrangian}
\eeq
These are all the terms which can contribute to the axial currents up to $\ord{\xi}$, which will determine the axial couplings up to $\ord{\xi^2}$, i.e., NNLO, which is what is needed for our purpose. There are several very important observations concerning the $\xi$-power counting. The corrections to the axial couplings start at $\ord{\xi}$,  and   the individual contributions of the different baryons in the loop diagrams are also $\ord{\xi}$.  Even  the difference of different axial couplings $ g_A^{BB'}- g_A^{B''B'''}$ starts at $\ord{\xi}$.  These  latter differences are UV finite.  The large $N_c$ cancellations do not seem manifest. However, at $N_c=3$,  where  the $\xi$-expansion is used,  cancellations do occur numerically  as  shown by Fig.~\ref{fig:gA-contrib} in Section V. Thus, the smallness of $\ord{\xi}$ terms in the axial couplings is a result of the incipient  manifestation of the cancellations in the large $N_c$ limit. If one would consider the strict large $N_c$ limit, the one loop corrections and counter-terms considered give the following $1/N_c$ power counting:
\bea
 g_A^{BB'}-\frac{5}{6}\mathring{g}_A
&=&\ord{M_\pi^2}+\log\left(\frac{M_\pi^2}{\mu^2}\right) \ord{\frac{M_\pi^2}{N_c}}+\cdots~~,\nonumber\\
  g_A^{BB'}- g_A^{B''B'''}&=&\ord{\frac{1}{N_c^2}}+\ord{\frac{M_\pi}{N_c^2}},
  \label{eq:gABB}
 \eea 
where, as expected, the latter differences are UV finite as in the $\xi$ expansion.

The explicit expression for $g_A^{NN}$ at $\ord{\xi^2}$ is give here for completeness:
\bea
g_A^{NN}&=& \frac{5}{6} \mathring{g}_A + \frac{5}{12 N_c^2} (3 C^A_2 + 2 N_c (C^A_0 + C^A_1 M_\pi^2 N_c)) \nonumber\\
&+& \frac{5 \mathring{g}_A^3 (4 + N_c)}{
 6 C_{HF} F_\pi^2 N_c^2(36 \pi)^2}
  \left\{-18 C_{HF}^3 - 12 C_{HF} M_\pi^2 N_c^2 - 
    9\pi C_{HF}^2 \sqrt{-9 C_{HF}^2 + M_\pi^2 N_c^2}\right.\nonumber\\
    & +& 
    2 \pi M_\pi^2 N_c^2 (M_\pi N_c - \sqrt{-9 C_{HF}^2 + M_\pi^2 N_c^2}\;)+ 
    27 C_{HF}^3 \log\frac{M_\pi^2}{\mu^2} \nonumber\\
    &+& \left.
    2 \sqrt{9 C_{HF}^2 - M_\pi^2 N_c^2} \;(9 C_{HF}^2 + 2 M_\pi^2 N_c^2) \tanh^{-1}\left(\frac{
      3 C_{HF}}{\sqrt{9 C_{HF}^2 - M_\pi^2 N_c^2}} \right)\right\}.
\eea
While in next section a discussion of the nucleon's $g_A$ in the context of  LQCD results is given, one can readily make an estimate of the spin-flavor symmetry breaking terms in the axial couplings $g_A^{NN}$ vs  $g_A^{\Delta N}$ using the result for the $\Delta$ width:
\bea
\Gamma_{\Delta\to \pi N}&=&\frac{1}{12 \pi  }\left(\frac 6 5 \frac{ g_A^{\Delta N}}{ F_\pi}\right)^2((m_\Delta-m_N)^2-M_\pi^2)^{3/2}.
\label{eq:Delta-width}
\eea
 Using the experimental value $\Gamma_{\Delta\to \pi N}(Exp)=116-120$ MeV \cite{Beringer:1900zz}, one obtains $g_A^{\Delta N}=1.235\pm 0.011$, which is remarkably  close to $g_A^{NN}= 1.2701 \pm 0.0025 $ \cite{Beringer:1900zz}.

\section{Analysis of lattice QCD results for baryon masses and the nucleon's axial coupling}
\label{sec:Lattice}
  
 As an application of the present framework of the $\xi$-expansion,  this section  presents an analysis of LQCD results for baryon masses and the nucleon's axial coupling.

Lattice QCD calculations of the non-strange ground state baryon masses (both of $N$ and $\Delta$ baryons) have opened the possibility of determining the quark mass dependencies, and similarly for the axial coupling of the nucleon.
These calculations represent a very fruitful ground of applications for ChPT, allowing in particular for a study of the convergence of the low energy expansion.
Current dynamical two- and  three-light-flavor   calculations of the hadron spectrum, and in particular of baryon masses,   with fixed strange quark mass and variable  $m_u=m_d$~ \cite{Durr:2008zz,WalkerLoud:2008bp,Aoki:2008sm,Lin:2008pr,Aoki:2009ix,Alexandrou:2009qu,Aoki:2010dy,Bietenholz:2011qq} 
are achieving remarkably accurate results in a range of quark masses where extrapolations to the physical limit are  now possible using effective theory.  
All calculations present similar results for the $N$ and $\Delta$ masses, namely, roughly linear dependencies of the masses  as a function of $M_\pi$, and extrapolations to the correct   physical  value  within a few percent. 
For the nucleon axial coupling
 $g_A^{NN}$  the results are  particularly interesting ~\cite{Edwards:2005ym,Bratt:2010jn,Alexandrou:2010hf,Yamazaki:2008py,Yamazaki:2009zq,Lin:2008uz} because they show small dependence  in a broad range of   $M_\pi$. The  most recent LQCD calculations for $N_f=2$~\cite{Edwards:2005ym,Lin:2008uz,Alexandrou:2010hf} and $N_f=2+1$~\cite{Yamazaki:2008py,Yamazaki:2009zq,Bratt:2010jn},   all agree on that observation. An open issue is that all  calculations give an underestimation for the value of $g_A^{NN}$ of about  $12\%$ below the experimental value.  
 
 Effects due to finite volume of the lattice have been studied for the observables considered here. Those effects are determined primarily by the value of  the product $L M_\pi$, where $L$ is the length of the lattice.
 For the baryon masses, the rule $L M_\pi   \gtrsim 4$~\cite{Alexandrou:2011iu} seems to be sufficient for the volume effects to be negligibly small.
On the other hand, for $g_A^{NN}$ the LQCD understanding of the finite volume effects is not yet complete. 
According to Ref.~\cite{Yamazaki:2008py}, $g_A^{NN}$ clearly exhibits  scaling
in $L M_\pi$ and in lattices with $L M_\pi \sim 4-5$ the effect on $g_A^{NN}$ is a 9 \% reduction in calculations with $2+1$ flavors of domain wall fermions and  a 25 \% reduction in calculations with two flavors of Wilson fermions. 
This has led to the current view that $L M_\pi \gtrsim  5 - 6$ or even higher may
in fact be needed to reliably determine $g_A^{NN}$.
Finite-volume effects for mases and the nucleon axial coupling have
been studied in effective theories~\cite{Beane:2004rf,Beane:2004tw,Colangelo:2005cg,Khan:2006de,Bedaque:2004dt,Detmold:2004ap,Smigielski:2007pe,Ishikawa:2009vc,Beane:2011pc,Geng:2011wq}. 
A detailed study of these effects in  the present formalism is beyond the scope
of this work, and  will be presented elsewhere~\cite{Alvaro-FVE}.

In the following,   combined fits to LQCD results for $N$ and $\Delta$ masses and the nucleon $g_A$  as functions of $M_\pi$ are carried out. 
For the $N$ and $\Delta$ masses  the results used are those from the PACS-CS collaboration of Ref.~\cite{Aoki:2008sm} and the LHP collaboration of Ref.~\cite{WalkerLoud:2008bp}.
For $g_A^{NN}$   the results used are those  from the LHP collaboration~\cite{Bratt:2010jn} and from the ETM collaboration~\cite{Alexandrou:2010hf}.  All collaborations obtain results satisfying the constraint    $L M_\pi  \gtrsim 4$ and for  quark masses  reaching down  close to the physical point, in particular for the baryon masses. The fits are carried out only including results where $L M_\pi  \gtrsim 4$.
The analysis of these LQCD results is carried out up to $\ord{\xi^3}$ for the masses and $\ord{\xi^2}$ for $g_A^{NN}$.
The  set of Lagrangian counter-terms is the one displayed in Eqs.~\eqref{eq:self-energy-CT-lagrangian} and~\eqref{eq:axial-current-CT-lagrangian}, which  are summarized by the following equations: 
\bea
\label{eq:dSigmaCT}
\delta \Sigma^{CT}(\p = 0) (S) &=&
  \frac{m_1}{N_c} 
+ \frac{C_{HF1}}{N_c^2} \vec S^2
+ \frac{C_{HF2}}{N_c^3} \vec S^4
+ \mu_1 M_\pi^2 + \frac{\mu_2}{N_c} \vec S^2 M_\pi^2 \, , \nonumber\\
\label{eq:dZCT}
\delta Z^{CT}(S) &=& 
  \frac{w_1}{N_c} 
+ \frac{w_2}{N_c} \vec S^2
+ \frac{w_3}{N_c^3}\vec{S}^4
+ z_1 N_c M_\pi^2 + \frac{z_2}{N_c} \vec S^2 M_\pi^2 \, , \nonumber\\
\label{eq:dgACT}
\delta g_A^{CT}(S,S') &=& \frac{C^A_0}{N_c} + C^A_1 M_\pi^2 + \frac{C^A_2}{N_c^2} (\vec S^2 +\vec S '^2 ) + \frac{C^A_3}{N_c^2} (\vec S^2 - \vec S '^2 ) \nonumber\\ &+ &  \frac{4 C^A_4}{N_c(2+N_c)}\delta_{SS'} \vec{S}^2 \, .
\eea

There are several LECs, which in order to be determined,  require knowledge of results at different values of $N_c$. With the  LQCD results at fixed $N_c=3$, those LECs combine with existing ones at lower order, making their determination impossible. Because   LQCD results on $g_A^{N\Delta}$ are not analyzed  and the lack of results for  $g_A^{\Delta\Delta}$, LECs which split the values of the different $g_A$'s cannot be fixed either.
For instance,  the LECs $m_1$ and $w_1$ give the sub-leading $N_c$ dependence  of $m_0$,  and therefore at fixed $N_c=3$ they are absorbed into the fitted value of $m_0$. The same will happen with $C_{HF1}$ and $w_2$ with $C_{HF}$. The LEC $\mu_1$ is a correction to the LO $\sigma$-term LEC $c_1$.
Similarly, one cannot separate  $C^A_0$ from $\mathring{g}_A$. Therefore, without loss of generality at fixed $N_c=3$,  the redundant LECs can be set to vanish. 
In addition, since the current fits only involve the nucleon's axial coupling, $g_A^{NN}$, not all LECs affecting the axial currents can be determined as mentioned earlier.
In particular,  counter-terms with commutators in Eq.~\eqref{eq:axial-current-CT-lagrangian}   only appear in  $g_A^{N\Delta}$.
Of course, depending on the order in the $\xi$-expansion, the number of LECs varies. Specifically, at leading order (LO), that is $\ord{\xi}$ for the mass and $\ord{\xi^0}$ for the axial coupling, the LECs are $m_0$, $\mathring{g}_A$, $C_{HF}$ and $c_1$, at NLO the additional LECs $C_{HF1}$, $\mu_1$ and $C^A_0$   appear.   Finally, at  NNLO, that is $\ord{\xi^3}$ for the mass and $\ord{\xi^2}$ for the axial coupling,  the additional LECs $\mu_2$, $z_1$ and $C^A_1$, which are fitted, and $\mu_3$, $w_1$, $w_2$ and $C^A_{2,3,4}$ that cannot be determined, make their appearance. 

\begin{figure}[h]
\begin{center}
\includegraphics[height=6cm,width=8cm,angle=0]{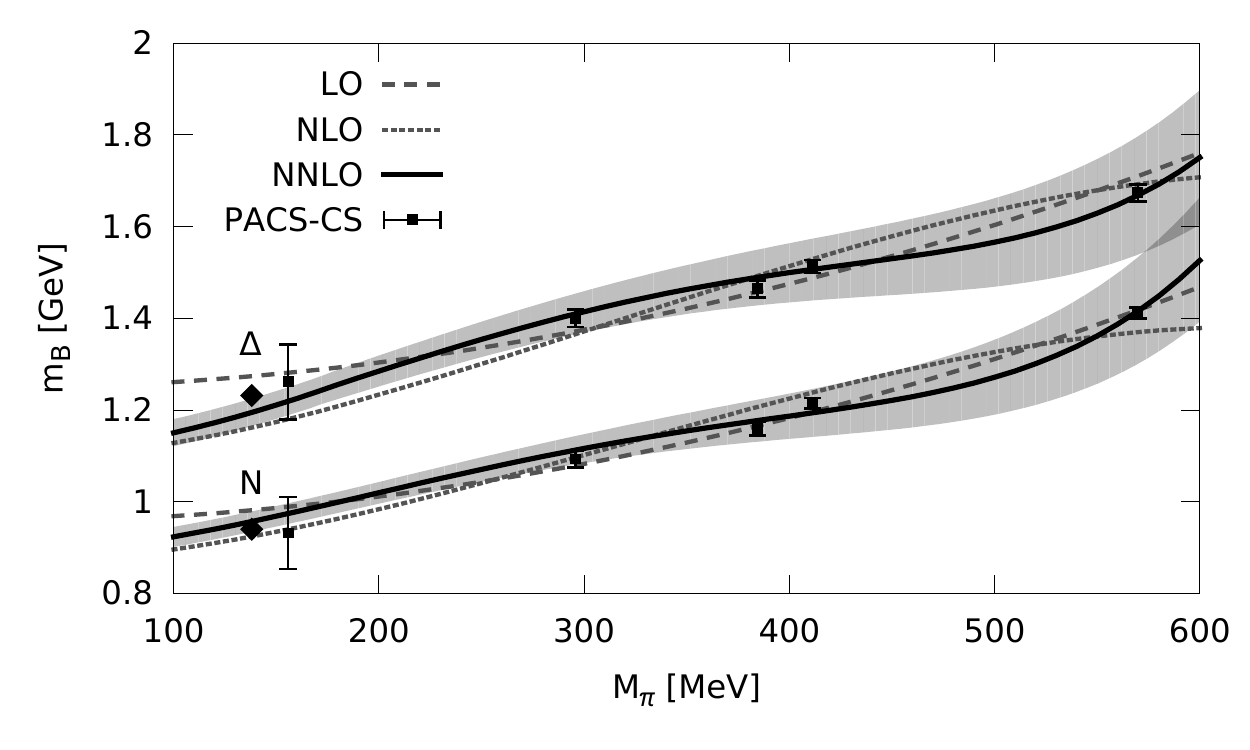}
\includegraphics[height=6cm,width=8cm,angle=0]{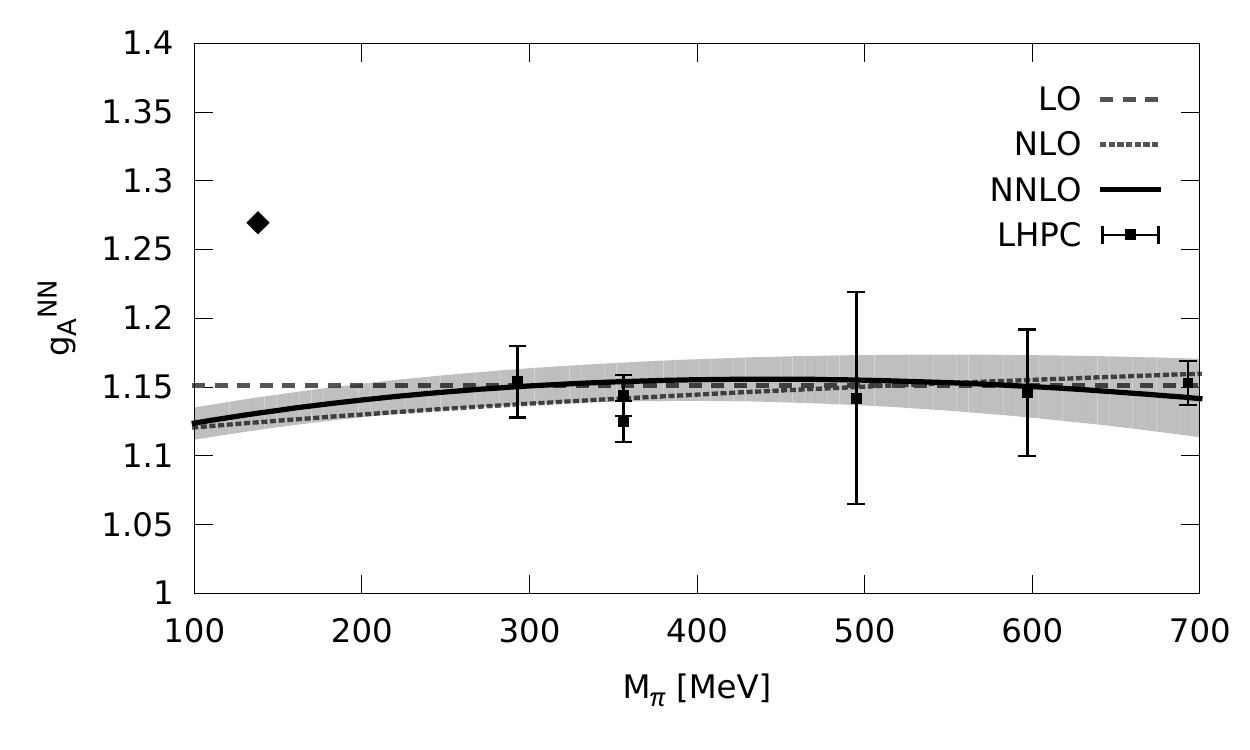}
\end{center}
\caption{Combined fits to PACS-CS~\cite{Aoki:2008sm} and LHP~\cite{Bratt:2010jn} corresponding to the results shown in the first row of Table~\ref{tab:fits}. The diamonds depict the physical values. The fits correspond to: LO (long-dashed line), NLO (short-dashed line) and NNLO (solid line). The bands correspond to the theoretical $68\%$ confidence interval.}
\label{fig:NNLO-fit}
\end{figure}

 The combined fits to $N$ and $\Delta$ masses and to $g_A^{NN}$ up to NNLO for the four possible combinations of  LQCD results from the collaborations considered here are presented in Table~\ref{tab:fits}, which shows the values for LECs obtained from the fits and   the extrapolated values for $m_N$, $m_\Delta$ and $g_A$ to the physical point. To estimate the theoretical errors,     the original lattice results are bootstrapped by Montecarlo, and the errors correspond to a 68\% confidence interval. 
 In the fits, for the masses the range $M_\pi<600$ MeV is used while for the axial coupling of the nucleon the range $M_\pi<700$ MeV is used. It is expected that the radius of convergence of the low energy expansion is smaller for the baryon masses than for $g_A$; this is because in the latter case the discussed cancellations reduce the $M_\pi$ dependence, while the lack of such cancellations for the loop contribution to the masses is magnified by $N_c$.
  The combined fits   are displayed in Fig \ref{fig:NNLO-fit}, which shows  the  LO to NNLO fits of LQCD results from the PACS-CS and LHP collaborations. 
\begin{centering}
\begin{sidewaystable}
\resizebox{1.0\textwidth}{!}
{
\begin{tabular}{|c|c|c|c|c|c|c|c|c|c|c|c|c|c|}
\hline\hline
LQCD Input & Order & $m_0$ [MeV] & $C_{HF}$ [MeV] & $\mathring{g}_A$ & $c_1$ [MeV$^{-1}$] & $\mu_2$ [MeV$^{-1}$] & $z_1$ ($10^{-7}$) [MeV$^{-2}$] & $C^A_1$ ($10^{-7}$) & $m_N$ [MeV] & $m_\Delta$ [MeV] & $g_A^{NN}$ & $\chi^2$& $\chi^2/DOF$ \\\hline
\multirow{2}{*}{PACS-CS $+$ LHP}
%
&  LO  & 294 (4) & 292 (10) & 1.38 (1) & 0.00048 (2) &  - & - & - & 981 (10) & 1274 (10) & 1.151 (8) &44    & 2.9\\
%
&  NLO & 262 (4) & 191 (6) & 1.38 (1) & 0.00153 (2) &  - & - & - & 925 (10) & 1162 (10) & 1.124 (7) & 74 & 4.9\\
%
&  NNLO  & 264 (7) & 170 (9) & 1.43 (2) & 0.00230 (7) & -0.0002 (1) & -9.0 (5) & -1.7 (6) & 957 (14) & 1195 (14) & 1.131 (12) & 40  & 3.3\\\hline
\multirow{2}{*}{LHP $+$ LHP}
%
&  LO  & 308 (3) & 335 (9) & 1.38 (1) & 0.000369 (12) & - & - & - & 1028 (8) & 1364 (10) & 1.151 (8) & 142  & 10.9\\
%
&  NLO  & 264 (3) & 216 (4) & 1.38 (1) & 0.00149 (2) & - & - & - & 944 (8) & 1217 (9) & 1.120 (8) & 145  & 11.2\\
%
&  NNLO  & 237 (6) & 212 (8) & 1.43 (2) & 0.00246 (6) & -0.00050 (9) & -5.0 (4) & -2.0 (6) & 924 (12) & 1219 (13) & 1.117 (11) & 56  & 5.6\\\hline
\multirow{2}{*}{PACS-CS $+$ ETM}
&  LO  & 294 (4) & 292 (10) & 1.389 (9) & 0.00048 (2) & - & - & - & 982 (11) & 1274 (11) & 1.157 (8) & 46 & 2.3\\
%
&  NLO & 262 (4) & 190 (5) & 1.387 (9) & 0.00155 (2) & - & - & - & 923 (10) & 1161 (10) & 1.131 (7) & 77 & 3.9\\
%
&  NNLO  & 263 (7) & 167 (8) & 1.47 (3) & 0.00241 (9) & -0.0002 (1) & -9.3 (5) & -4 (2) & 956 (14) & 1192 (14) & 1.16 (2) & 40 & 2.4\\\hline
\multirow{2}{*}{LHP $+$ ETM}
&  LO  & 308 (3) & 335 (9) & 1.39 (1) & 0.00037 (1) &  - & - & - & 1029 (7) & 1364 (9) & 1.157 (9) & 146 & 8.1\\
%
&  NLO  & 264 (3) & 215 (4) & 1.39 (1) & 0.00150 (2) & - & - & - & 942 (8) & 1215 (8) & 1.127 (8) & 149 & 8.3\\
%
&  NNLO  & 235 (6) & 209 (8) & 1.45 (2) & 0.00254 (8) & -0.00051 (8) & -5.1 (4) & -3.5 (1.4) & 921 (12) & 1214 (13) & 1.134 (15) & 58 & 3.9\\
\hline \hline
 \multicolumn{1}{| c}{Natural magnitude of LECs}  &&$\sim 300$&$\sim 300$&1-2&$10^{-3}$&$10^{-3}$&$10^{-6}$&$10^{-6}$\\\cline{1-9}\cline{1-9}
\end{tabular}
}
\caption{Results for LECs and extrapolated masses $m_N$ and $Re(m_\Delta)$,  and the nucleon $g_A$ from a combined fit to LQCD. The renormalization scale has been taken to be $\mu=700$ MeV. Lattice results for masses are from the PACS-CS~\cite{Aoki:2008sm} and  LHP~\cite{WalkerLoud:2008bp} collaborations, and for $g_A^{NN}$ from LHP~\cite{WalkerLoud:2008bp,Bratt:2010jn}  and ETM~\cite{Alexandrou:2010hf} collaborations. Different orders in the $\xi$-expansion are shown. The LECs not shown have been set to vanish. { Remark: the physical values of the baryon masses and $g_A^{NN}$ have been assigned errors of the same order as the theoretical error in the effective theory: $\sim \pm 20$ MeV for the $N$ and $\Delta$ masses, and $\sim \pm .03$ for $g_A^{NN}$.}}
\label{tab:fits}
\end{sidewaystable}
\end{centering}
\begin{figure}[h]
\begin{center}
\includegraphics[height=6cm,width=8cm,angle=0]{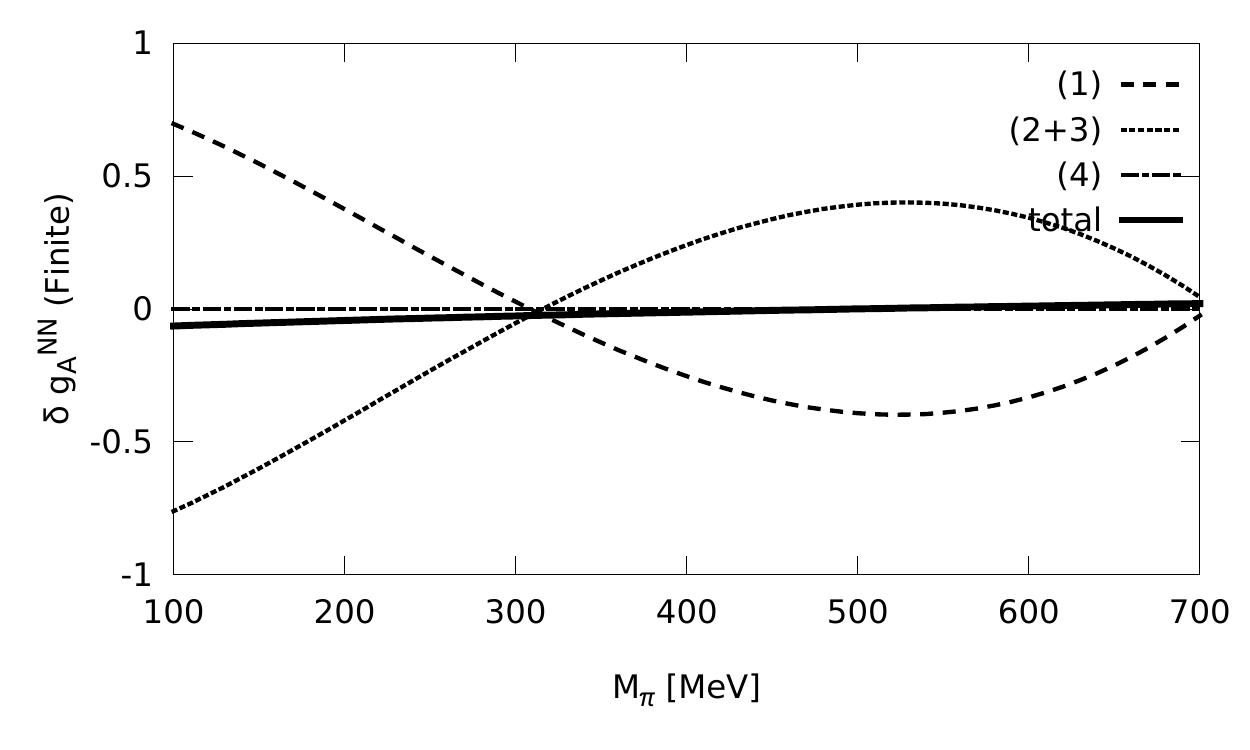}
\includegraphics[height=6cm,width=8cm,angle=0]{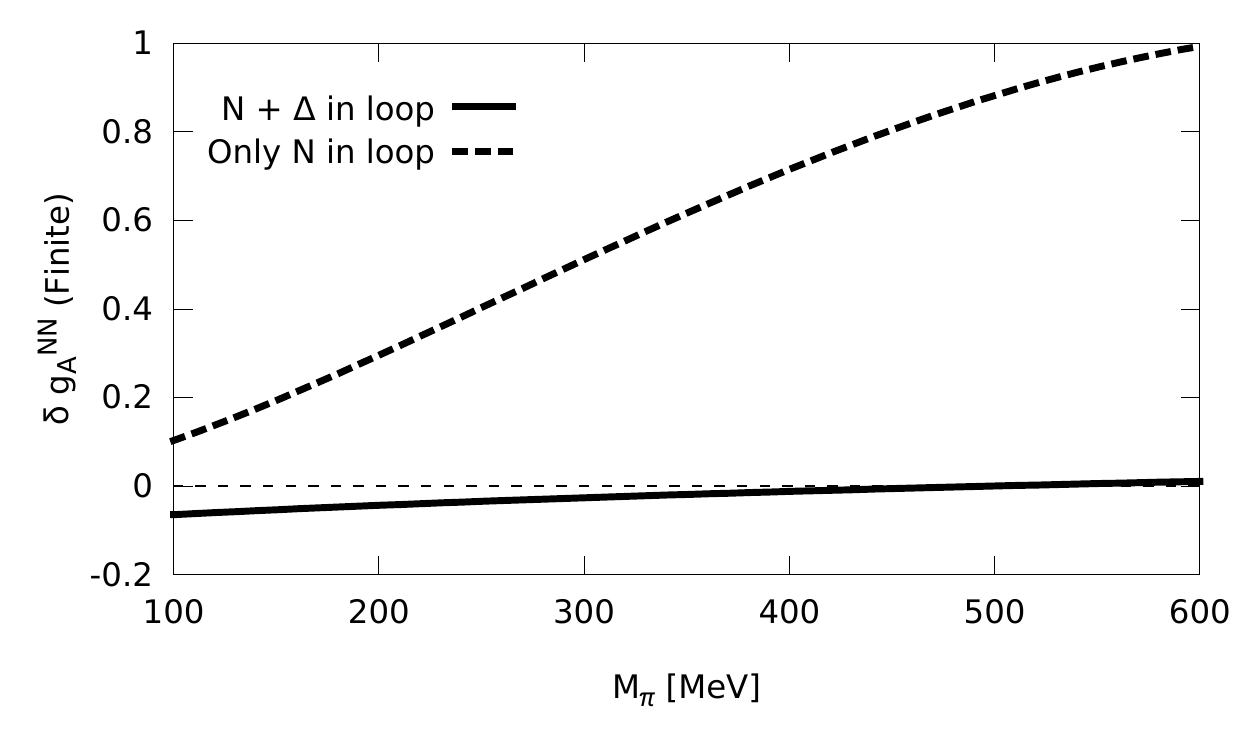}\\
\includegraphics[height=6cm,width=8cm,angle=0]{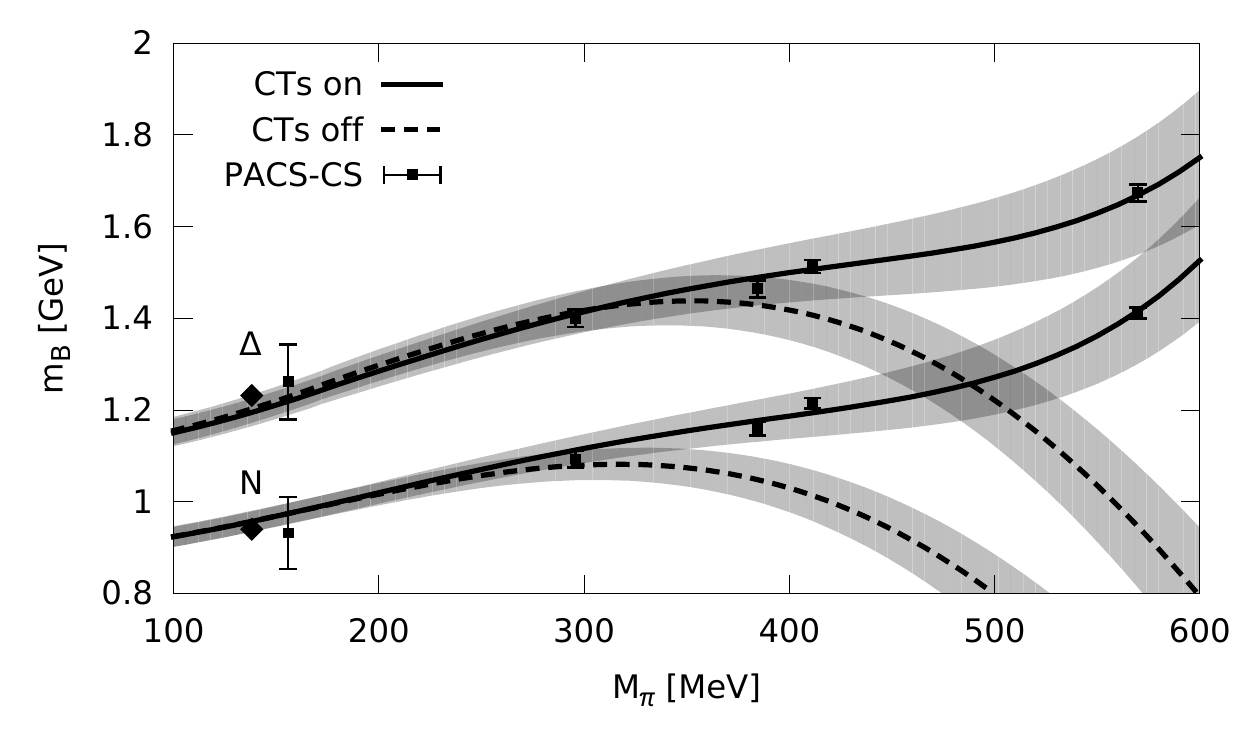}
\end{center}
\caption{Finite parts of the one-loop contributions to $g_A^{NN}$: the upper left panel shows the individual contributions of the diagrams in Fig.~2
up to $\ord{\xi^3}$, and
 the right panel shows the   effect of switching off the contribution of the  $\Delta$ in the loops. The third panel shows the effect of removing the contributions of the counter-terms to the masses.  Throughout $\mu=700$ MeV.}
\label{fig:gA-contrib}
\end{figure}

The following remarks on the fits are in order:
\begin{enumerate}
\item All fitted  LECs are of natural size when the renormalization scale is taken to be $\mu\sim m_\rho$.  
\item Parameters appearing at lower orders, namely $m_0,~\mathring{g}_A$ and $C_{HF}$,  remain stable at higher orders, except   $c_1$ that  changes by more than the estimated 30\% when increasing the order in $\xi$  of the fit by one unit.
\item For baryon masses, LQCD data and physical point values are consistent even at LO,  where with  only  three parameters one can extrapolate to the physical values and get a good fit up to $M_\pi\sim 350$ MeV as shown in Fig.~\ref{fig:NNLO-fit}.   For larger values of $M_\pi$  an approximate linear fit  is  consistent~\cite{WalkerLoud:2008bp} in the range $M_\pi^{phys}<M_\pi<450$ MeV. Since at LO there are contributions to the baryon masses which are  proportional to $  N_c \, c_1 M_\pi^2$, the NLO and NNLO effects are necessary to give the approximate linear behavior in that range of $M_\pi$.
\item For the case of the axial current, cancellations of large contributions from individual loop diagrams   are very pronounced and the almost flat behavior of $g_A^{NN}$  as a function of $M_\pi$ obtained in LQCD is naturally explained. This is shown in the upper left panel of  Fig.~\ref{fig:gA-contrib} which depicts the finite one-loop contributions to $g^{NN}_A$ from each  diagram ($\mu=700$ MeV). As stated in Eq.~\eqref{eq:gA-cancel} this cancellation is exact in the large $N_c$ limit. However, at $N_c=3$ this cancellation is not exact  but  still quite pronounced (solid curve in upper left panel of Fig.~\ref{fig:gA-contrib}), and plays the key role in explaining the small dependence in $M_\pi$. 
A similar cancellation occurs between the contributions of $N$ and $\Delta$ in the loop contributions. This is shown in the upper right panel of  Fig.~\ref{fig:gA-contrib}.
\item The physical $g_A^{NN}$ cannot be fitted along with the lattice results, instead the lattice results and the expansion to NNLO extrapolate to a value 12\% smaller  than the physical one, as clearly shown in Fig. \ref{fig:NNLO-fit}. The recent LQCD results \cite{Green:2012ud} which reach further down in $M_\pi$ continue that trend. On the other hand, recent LQCD results for $g_A^{NN}$ from the CLS collaboration 
\cite{Capitani:2012gj} can be made compatible with the physical value, but the error bars for $M_\pi<400$ MeV are quite large, and thus they cannot be considered to be significantly different than the ones of the LHP collaboration depicted in Fig. \ref{fig:NNLO-fit}.  It seems therefore, that the LQCD calculations are still evolving and it is possible that soon the origin of the  mentioned discrepancy will be elucidated.

The argument that the current LQCD results are correct and that the failure to extrapolate to the correct physical value is a problem of the effective theory seems unlikely on the following grounds.   It is evident from  Fig. \ref{fig:NNLO-fit} that in that case the effective theory should give up to a 12\% enhancement below $M_\pi<300$ MeV. Since that  does not occur at the order calculated here, namely NNLO, it should be provided by NNNLO contributions. The latter contributions are $\ord{\xi^3}$, and estimating that the effective value of the expansion parameter $\xi$ in the
mass range of the physical pion mass is ~1/3 to 1/4, one concludes that NNNLO corrections cannot be larger than a few percent.
 
%
\item A fit restricted only to masses gives too small a value for $g_A^{NN}$, namely, $g_A^{NN}\sim 0.5-0.8$. A realistic value can only be obtained with the combined fit.
\item Predictions for $g_A^{N\Delta}$ and $g_A^{\Delta\Delta}$ cannot be made without the corresponding LQCD results. However, the results at the physical point from Eq.~(\ref{eq:Delta-width}) suggest that these are going to be very similar in value to $g_A^{NN}$. Further efforts to study these couplings in LQCD will be very useful.
\item In the masses one finds that above $M_\pi>350-400$ MeV there is a significant cancellation between the contributions of the one-loop diagram  and the counter-terms  as shown in   Fig~\ref{fig:gA-contrib}, which   must be taken as an indicator of the range of convergence of the expansion.  Note that the   mass counter-terms are $\ord{M_\pi^2 N_c^0}=\ord{\xi^2}$. 
\item Evaluating the   $\sigma$ terms at the physical pion mass using   the fits in Table~\ref{tab:fits},  the results  shown in Table~\ref{tab:sigma_terms} are obtained.

\begin{centering}
\begin{table*}[ttt]
{
\begin{tabular}{|c|c|c|c|}
\hline\hline
LQCD Input  & Order & $\sigma_{ N}$ [MeV] & $\sigma_{\Delta}$ [MeV] \\\hline
\multirow{2}{*}{PACS-CS $+$ LHP}
&  LO   & 27   & 27  \\
&  NLO  & 58  & 68  \\
&  NNLO & 66 (4)     & 90 (5)    \\
\hline
\multirow{2}{*}{LHP $+$ LHP}
&  LO   & 21  & 21  \\ 
&  NLO  & 55   & 66  \\
&  NNLO & 76  (4)    & 99 (4) \\
\hline
\hline 
\end{tabular}
}
\caption{Results for the $N$ and $\Delta$  $\sigma$-terms. These results  correspond to the fits in the first two rows of Table~\ref{tab:fits}.  }
\label{tab:sigma_terms}
\end{table*}
\end{centering}

%
%
 It is evident from the important change in the results from NLO to NNLO that the $\sigma$ terms cannot yet be very accurately determined from the    current LQCD results. One finds that the   $\sigma$ terms  do not depend significantly on the choice of LQCD results for $g_A^{NN}$.  $\sigma$ terms were obtained in other analyses of LQCD results in the framework of $SU(3)$ BChPT with $\Delta$ included in Ref. \cite{MartinCamalich:2010fp}. The present results at NLO are compatible with theirs, but are substantially larger at NNLO. However, if the fit is required to pass through the physical baryon masses, for $\sigma_N$ the NNLO is similar to that in \cite{MartinCamalich:2010fp}, however,   the result obtained here where $\sigma_N<\sigma_\Delta$, is opposite to the one in  \cite{MartinCamalich:2010fp}. This indicates that the $\sigma$ terms are   sensitive to the particular formulation of the effective theory and also to the order of the expansion, an issue which remains to be clarified.
\item It must be emphasized that the results obtained here have many similarities with  those obtained in works where the $\Delta$ has been included explicitly \cite{Procura:2003ig,Bernard:2005fy,Procura:2006bj,Procura:2006gq,Bernard:2007zu,MartinCamalich:2010fp,Semke:2011ez}.  The main advantage of the present approach of the $\xi$-expansion is its systematic character,  which in particular will be more prominently   shown when carrying out  higher order calculations  than the ones considered here.
\end{enumerate}

\section{Discussion and conclusions}
\label{sec:Conclusions}

Chiral symmetry and the large $N_c$ limit  are of fundamental conceptual importance in QCD. The former is known to play a crucial role in light hadrons, and there are multiple indications that the latter is also important,    in particular for baryons.  It is therefore very important to have a theoretical  framework where both  of these aspects of QCD are consistently incorporated. 
This is possible with the combined $1/N_c$ and Chiral expansions  of QCD, which in the baryon  sector is implemented with the effective theory discussed in this work. A particular power counting, the $\xi$-expansion, which links the $1/N_c$ and low energy expansions as $1/N_c=\ord{\xi}=\ord{p}$ is proposed as the most realistic one for studying baryons at $N_c=3$. Results for the masses and axial couplings at NNLO have been given, and applied to current LQCD results.

The $\xi$-expansion at NNLO clearly provides a satisfactory description of the LQCD results, and in particular it  illuminates the  mild dependence of the axial couplings on the quark masses as a result of important cancellations, which had been realized in various previous analysis by various groups.  It is important to complete the study in $SU(3)$, in particular because the one-loop  contributions to the baryon masses  become larger in magnitude, and a smaller range of convergence is expected \cite{Semke:2011ez}. 
These results will be presented elsewhere  \cite{SU3:AlvaroJose}. Recently, results for the axial currents with three flavors in a similar framework  to the one developed here were presented in Ref.  \cite{FloresMendieta:2012dn}.

 The deficit in $g_A^{NN}$ at the physical point is expected to be a LQCD issue rather than a problem of convergence of the effective theory.
The main reason for this expectation is that    the $\xi$ expansion is especially well behaved for $g_A$.  Among the possible sources of systematic errors in the extraction of $g_A$ from LQCD calculations  might be the finite volume effects and/or  the contamination in the three-point functions by excited baryon states. 

In addition to the tests LQCD can provide on   quark mass dependencies, it is also an ideal tool to test the $N_c$ behavior of QCD. Baryon LQCD is becoming accessible at varying values of $N_c$ \cite{DeGrand:2012hd}, which is a promising development.

\begin{acknowledgments}
The authors thank C. Alexandrou, K. Kanaya and D.  Renner for useful discussions on several aspects of the LQCD results.
This work was supported by DOE Contract No. DE-AC05-06OR23177 under which JSA operates the Thomas Jefferson National Accelerator Facility, and by the National Science Foundation (USA) through grant PHY-0855789 (JLG). 
\end{acknowledgments}
\newpage

\appendix

\section{Spin-flavor Algebra}
\label{sec:Algebra}

The $4N_f^2-1$ generators of the spin-flavor group $SU(2 N_f)$  consist of the three spin generators $S^i$, the $N_f^2-1$   flavor $SU(N_f)$ generators  $T^a$, and the remaining $3(N_f^2-1)$   spin/flavor generators $G^{ia}$. The commutation relations are:
\bea
~& [ S^i,S^j ]= i \eps_{ijk}S^k,~~ [ T^a,T^b ]=i f_{abc} T^c ,~~ [ T^a,S^i ]=0& ,\nonumber\\    
~& [ S^i,G^{ja} ]=i  \eps_{ijk} G^{ka},~~ [T^a,G^{ib}]=i f_{abc} G^{ic}&,\nonumber\\   
~ & [ G^{ia},G^{jb} ]= \frac{i}{4}\delta^{ij}f^{abc}T^c+\frac{i}{2 N_f}\delta^{ab} \eps^{ijk} S^k+\frac{i}{2}\eps^{ijk}d^{abc}G^{kc}&.
\label{eq:commutation-relations}
\eea
For two flavors  one has the isospin generators  $I^a$ $a=1,2,3$.

In representations with $N_c$ indices (baryons), the generators $G^{ia}$ have matrix elements $\ord{N_c}$ on states with $S=\ord{N_c^0}$. A contracted  $SU(4)$ algebra  is  defined by the generators $\{S^i,I^a,X^{ia}\}$, where $X^{ia}=G^{ia}/N_c$.  In large $N_c$, the generators $X^{ia}$ become semiclassical as $[X^{ia},X^{jb}]=\ord{1/N_c^2}$, while having matrix elements $\ord{1}$ in  baryon representations.

\section{Non-linear realization of chiral symmetry and spin-flavor transformations}
\label{sec:Symmetries}

In the symmetric representations of $SU(4)$  the baryon spin-flavor multiplet consists of the baryon states with $I=S$. In particular,   isospin transformations   will act on the spin-flavor multiplet in an obvious way. This permits a straightforward implementation of the non-linear realization of chiral $SU_L(2)\times SU_R(2)$ on the spin-flavor multiplet. Defining as usual the Goldstone Boson fields $\pi^a$ through the unitary parametrization $u=\exp(i\frac{\pi^a I^a}{F_\pi})$ (note that in the fundamental representation $I^a=\tau^a/2$), for any isospin representation one defines a non-linear realization of chiral symmetry according to \cite{Coleman:1969sm,Callan:1969sn}:
\beq
(L,R):u=u'=R u h^\dag(L,R,u)=h(L,R,u)uL^\dag,
\label{eq:transformation-u}
\eeq
where $(L,R)$ is a   $SU_L(2)\times SU_R(2)$ transformation. This equation  defines $h$, and  since $h$ is an isospin $SU(2)$ transformation itself, it can be written as $h=\exp(i c^a I^a)$.  The chiral  transformation on the baryon multiplet $\B$ is then given by:
\beq
(L,R):\B=\B'=h(L,R,u)\B.
\label{eq:transformation-B}
\eeq
On the other hand, spin-flavor transformations of interest are the contracted ones, namely those generated by $\{S^i,I^a,X^{ia}=\frac{1}{N_c}G^{ia}\}$.
While the isospin transformations act on the pion fields in the usual way, and the spin transformations must be performed along with the corresponding spatial rotations.   The transformations generated by $X^{ia}$ are   defined  to only act on the baryons.

\section{Tools for building effective Lagrangians}
\label{sec:Tools}

The effective baryon Lagrangian can be expressed in the usual way as a series of terms which are $SU_L(2)\times SU_R(2)$ invariant (upon introduction of appropriate sources; see for instance \cite{Scherer:2002tk} for details).
In addition, implemented in the effective Lagrangian is the approximate $SU(4)$ symmetry and its breaking as a power series in $1/N_c$ \cite{Jenkins:1995gc}.
The fields in the effective Lagrangian are the Goldstone Bosons parametrized by the unitary $SU(2)$ matrix field $u$ and the baryons   given by  the symmetric $SU(4)$ multiplet   $\B$ of $I=S$ fields.  

The building blocks for the effective theory consist of low energy operators, and spin-flavor operators. 

The low energy operators are the usual ones, namely:
\bea
 D_\mu&=&\partial_\mu-i \Gamma_\mu,~~~\Gamma_\mu=\Gamma_\mu^\dag=\frac{1}{2}(u^\dag(i\partial_\mu+r_\mu)u+u(i\partial_\mu+\ell_\mu)u^\dag),\nonumber\\
u_\mu&=&u^\dag_\mu=u^\dag(i\partial_\mu+r_\mu)u	-u(i\partial_\mu+\ell_\mu)u^\dag,	\nonumber\\	
\chi&=&2 B_0(s+i p)	,~~
\chi_{\pm}=u^\dag  \chi u^\dag \pm u\chi^\dag  u,\nonumber\\
F^{\mu\nu}_{L}&=&\partial^\mu\ell^\nu-\partial^\nu\ell^\mu-i[\ell^\mu,\ell^\nu],~~F^{\mu\nu}_{R}=\partial^\mu r^\nu-\partial^\nu r^\mu-i[r^\mu,r^\nu],
\label{eq:chiral-building-blocks}
\eea
where $D_\mu$ is the chiral covariant derivative,  $s$ and $p$ are scalar and pseudo-scalar sources,  $\chi_{\pm}=2 M_\pi^2+\cdots$,  and $\ell_\mu$ and $r_\mu$ are gauge sources.
The spin-flavor operators are tensor operators consisting of products of the spin-flavor generators. These operators can be reduced by means of the commutation relations to forms which only contain anti-commutators. A   set of identities shown in Table \ref{RedRules} permits one to arrive at sets of basis operators at each order in $1/N_c$  for a given spin/isospin tensor type of operator. The $1/N_c$ order $\nu_O$ of an operator $O$, reduced as mentioned,  is   $\nu_O=n-1-\kappa$ \cite{Dashen:1994qi}, where $n$ is the number of generators appearing as factors in the operator (one then says that the operator is an $n$-body operator), and $\kappa$ is the number of  generators $G^{ia}$ in the product.

The leading order  equations of motion can be used in the construction of the higher order terms, namely, $iD_0\B= (\frac{C_{HF}}{N_c} S(S+1)+\frac{c_1}{2} N_c  \chi_+)\B$, and $\nabla_\mu u^\mu=\frac{i}{2}\chi_-$.

\section{Matrix elements of spin-flavor operators in the symmetric representations of ${\mathbf{SU(4)}}$}
\label{sec:ME}

The evaluation of the matrix elements of spin-flavor operators in the present work can be carried out starting from the following matrix elements of the spin-flavor generators in the totally symmetric representation of $SU(4)$ corresponding to the Young tableux with a single row of $N_c$ boxes. The basis states of the symmetric representation consists of the states with $I=S$, namely $\mid S\,S_3I_3\rangle$, where $S_3$ and $I_3$ are the spin and isospin projections respectively.
{\small
\bea
\langle  S'\,S'_3I'_3  \mid  S^i \mid  S\,S_3I_3 \rangle&=&\sqrt{S(S+1)} \delta_{SS'} \delta_{I_3 I'_3} \langle S\,S_3,1 i \mid S'\,S'_3\rangle,\nonumber\\
\langle  S'\,S'_3I'_3  \mid I^a  \mid  S\,S_3I_3 \rangle&=&\sqrt{S(S+1)}    \delta_{SS'} \delta_{S_3 S'_3}      \langle  S\,I_3,1a\mid S' I'_3\rangle,\nonumber\\
\langle  S'\,S'_3I'_3  \mid  G^{ia} \mid  S\,S_3I_3 \rangle&=&\frac{1}{4}\sqrt{\frac{2S+1}{2S'+1}} \zeta(N_c,S,S')  \langle S\,S_3,1 i \mid S'\,S'_3\rangle    \langle  S\,I_3,1a\mid S' I'_3\rangle, 
\label{eq:generator-ME}
\eea}
where $\zeta(N_c,S,S')=\sqrt{(2+N_c)^2-(S-S')^2(S+S'+1)^2}$ \cite{Carlson:1998vx}.
The products of generators can be reduced by means of the use of the commutation relations, and further, for matrix elements in the symmetric representation, via the reduction rules \cite{Dashen:1994qi}, which  for convenience are displayed in   Table~\ref{RedRules}.
\begin{table}[h!]
\caption{\label{RedRules} $SU(4)$ operator identities    in the totally symmetric irreducible representation $(N_c,0,0)$ of $SU(4)$. The last column gives the operator's quantum numbers $(J,I)$ under $SU(2)\times SU(2) $}\vspace*{2mm}
\begin{tabular}{c|c}\hline\hline
$\{S^i,S^i\}-\{I^a,I^a\}=0$ &(0,0)\\
$\{S^i,S^i\}+\{I^a,I^a\}+4\{G^{ia},G^{ia}\}=\frac{3}{2}N_c(4+N_c)$&(0,0)\\
$2\{S^i,G^{ia}\}=(2+N_c) I^a $&(0,1)\\
$2\{I^a,G^{ia}\}=(2+N_c)  S^i $&(1,0)\\
$\frac{1}{2}\{S^k,I^c\}-\eps^{ijk}\eps^{abc}\{G^{ia},G^{jb}\}=(2+N_c)G^{kc}$ &(1,1)\\
$\eps^{ijk}\{S^i,G^{jc}\}=\eps^{abc}\{I^a,G^{kb}\}$&(1,1)\\
$4\{G^{ia},G^{ib}\}\arrowvert_{I=2}=\{I^a,I^b\}\arrowvert_{I=2} $&(0,2)\\
$4\{G^{ia},G^{ja}\}\arrowvert_{J=2}=\{S^i,S^j\}\arrowvert_{J=2} $&(2,0)\\ 
\hline\hline
 \end{tabular}
 \label{tab:redrel}
 \end{table}\\
 
{\bf Useful matrix elements:}

It is always convenient to express matrix elements in terms of reduced matrix elements (RMEs) defined in the ordinary Wigner-Eckart fashion \cite{Edmonds:1955fi}. The RMEs defined here are with respect to $SU_{\rm spin}(2)\times SU_I(2)$.  For matrix elements in the symmetric representation of spin-flavor the Wigner-Eckart theorem reads:
\beq
\langle  S'\,S'_3I'_3  \mid  {\mathbf O}^{\mathbf {JJ_3}}_{\mathbf {II_3} }\mid  S\,S_3I_3 \rangle=\frac{\langle S'\mid \mid  {\mathbf O}^{\mathbf {J}}_{\mathbf {I} }\mid \mid S\rangle}{2S'+1} \langle SS_3,{\mathbf{J J_3}} \mid S' S'_3 \rangle \langle S I_3, {\mathbf{I I_3}} \mid S'I'_3\rangle,
\label{eq:RME-def}
\eeq
where $ {\mathbf O}$ is an $SU_S(2)\times SU_I(2)$ irreducible tensor operator, and $\langle S'\mid \mid  {\mathbf O}^{\mathbf {J}}_{\mathbf {I} }\mid \mid S\rangle$ is the reduced matrix element. Note that the notation $\mid\mid S\rangle$ indicates the spin-flavor states in the symmetric representation $(N_c,0,0)$ with $I=S$.
The reduced matrix elements of the $SU(4)$ generators read:
\bea
\langle S'\mid \mid  S \mid \mid S\rangle&=&\delta_{SS'} (2S+1)\sqrt{S(S+1)},\nonumber\\
\langle S'\mid \mid I \mid \mid S\rangle&=&\delta_{SS'} (2S+1)\sqrt{S(S+1)},\nonumber\\
\langle S'\mid \mid G  \mid \mid S\rangle&=&\delta_{\{S,S',1\}}\frac{1}{4}\sqrt{(2S+1) (2S'+1)}\zeta(N_c,S,S'),
\label{eq:generator-RME}
\eea
where $\delta_{\{S,S',1\}}=1$ if $\mid S-S'\mid\leq1$ and otherwise vanishes.

Reduced matrix elements of the operators involving the projects ${\cal{P}}_n$ are easily obtained using that  ${\cal{P}}_n=\sum_{S_{n3},I_{n3}} \mid S_n, S_{n3}I_{n3}\rangle \langle S_n, S_{n3}I_{n3}\mid$, and the $SU(2)$  re-coupling results \cite{Edmonds:1955fi}. For the masses the relevant such RME becomes:
\beq
\langle S\mid\mid G^{ia}{\cal{P}}_nG^{ia}\mid\mid S\rangle= \frac{1}{2S+1}\langle S_n\mid\mid G\mid\mid S\rangle^2.
\label{eq:RME-GPG}
\eeq
For the axial currents the following RME is needed, namely:
\bea
\langle S'\mid\mid G^{jb}{\cal{P}}_{n'}G^{ia}{\cal{P}}_n G^{jb}\mid\mid S\rangle&=&\left\{\begin{array}{ccc}
S&S_n&1\\
S_{n'}&S'&1
\end{array}\right\}^2 \nonumber\\&\times&\langle S'\mid\mid G\mid\mid S_{n'}\rangle \langle S_{n'}\mid\mid G\mid\mid S_n\rangle \langle S_n\mid\mid G\mid\mid S\rangle.
\label{eq:RME-GPGPG}
\eea

Various reduced matrix elements which appear in the evaluation of the UV divergent pieces of the one-loop contributions to the self energy and to the axial currents are given below. They are obtained using the results given in the above Eqs.~\eqref{eq:RME-GPG} and \eqref{eq:RME-GPGPG}:

\beq
\langle S'\mid \mid  G^2\mid \mid S\rangle=\frac{1}{2}\delta_{SS'}(2S+1) ( -S (S + 1) + \frac{3}{8} N_c (4+N_c )),
\label{eq:RME-G2}
\eeq
where $G^2=G^{ia}G^{ia}$.
\beq
\langle S'\mid \mid  G \vec{S}^2 G\mid \mid S\rangle=\frac{1}{2}\delta_{SS'}(2S+1)\left(\frac{3}{4} N_c (4 + N_c) + \frac{1}{8} (-40 + 12 N_c + 3 N_c^2) S(S+1) - (S(S+1))^2\right),
\label{eq:RME-GS2G}
\eeq
\bea
\langle S'\mid \mid  G \vec{S}^4 G\mid \mid S\rangle&=&\frac{1}{2}\delta_{SS'}(2S+1)\left(\frac{3}{4} N_c (4 + N_c) + (-6 + 5 N_c + \frac{5}{4} N_c^2) S(S+1)\right.
\nonumber\\
& +&\left.  (-7 + 
    \frac{3}{4} N_c + \frac{3}{16} N_c^2) (S(S+1))
^2 - \frac{1}{2}(S(S+1))^3\right).
\label{eq:RME-GS4G}
\eea

In the following, $S'=S$ or $ S\pm 1$. With obvious notation:
\beq
\langle S'\mid \mid  \{S^i, G^{ia}\} \mid \mid S\rangle = \delta_{SS'} (1+\frac{N_c}{2})(2 S+1)\sqrt{ S (S+1 )}  ,
\eeq
obtained using the corresponding  reduction relation in Table~\ref{tab:redrel}.
\beq
\langle S'\mid \mid  S^i I^a \mid \mid S\rangle = \delta_{SS'} (2 S+1) S (S+1 )  ,
\eeq
\beq
\langle S'\mid \mid  S G^{ia} S\mid \mid S\rangle = \frac{1}{2}(S(S+1)+S'(S'+1)-2) \langle S'\mid \mid G  \mid \mid S\rangle,
\eeq
\bea
\langle S'\mid \mid  G G^{ia} G\mid \mid S\rangle &=& \frac{1}{16} \langle S'\mid \mid G  \mid \mid S\rangle\nonumber
\\
&\times &\left(3N_c(4+N_c)-4 (2+S(S+1)+S'(S'+1))\right),
\label{eq:RME-GGG}
\eea
\bea
\langle S'\mid \mid  G G^{ia} \vec{S}^2 G\mid \mid S\rangle&=&\frac{1}{2}\langle S'\mid \mid G  \mid \mid S\rangle\left(\frac{3}{4} N_c (4 + N_c)\right. \nonumber
\\
&+&\left.
S(S+1)( - S(S+1)
 -5 + \frac{3}{8} N_c (4 + N_c) + \frac{1}{2}S'(S'+1))\right.\nonumber\\
  &-&\left. S'(S'+1)(1+
  \frac{1}{2}S'(S'+1))
\right),
\label{eq:RME-GGS2G}
 \eea
\bea
\langle S'\mid \mid  G G^{ia} \vec{S}^4 G\mid \mid S\rangle&=&\frac{1}{2}\langle S'\mid \mid G  \mid \mid S\rangle \nonumber
\\
&\times&
\left( \frac{3}{2}N_c (4 + N_c) + (S(S+1))^2
 (-16 + \frac{3}{8} N_c
 (4 + N_c
) - 3 S'(S'+1)
) \right.\nonumber\\
&+&\left. (S'(S'+1))^2(1
 - \frac{3}{2}
  S'(S'+1)
) \right.\nonumber\\
&+& \left. S(S+1)
 (-12 + \frac{5}{2} N_c
 (4 + N_c
) + \frac{7}{2} (S'(S'+1))^2)
\right),
\label{eq:RME-GGS4G}
 \eea
\bea
\langle S'\mid \mid  G \vec{S}^2 G^{ia} \vec{S}^2 G\mid \mid S\rangle&=&\frac{1}{4}\langle S'\mid \mid G  \mid \mid S\rangle \nonumber
\\
&\times&\left(
 (2 + N_c)^2+\frac{1}{2}(-16 + 5 N_c (4 + N_c)) (S(S+1)+S'(S'+1)) \right.\nonumber\\
 &-&  (S(S+1))^2
 (9+ S(S+1))
-   (S'(S'+1))^2
 (9+ S'(S'+1))\nonumber\\
&  + &\left.
 \frac{3}{4} (-16 + N_c (4 + N_c))  S(S+1)S'(S'+1)\right).
\label{eq:RME-GS2GS2G}
 \eea
Finally, using that for any spin and isospin singlet operator (not necessarily an $SU(4)$ singlet) ${\cal{Q}}$, $\langle S'\mid \mid {\cal{Q}} {\mathbf{O^J_I}} \mid \mid S\rangle=\frac{1}{2 S'+1}\langle S'\mid \mid {\cal{Q}}\mid \mid S'\rangle \langle S'\mid \mid {\mathbf{O^J_I}}\mid \mid S\rangle$, one can easily obtain the rest of the matrix elements involved in the calculation of the axial currents.

\bibliography{Refs}

\end{document}